\documentclass[11pt,a4paper]{article}
\usepackage{jheppub}
\usepackage{physics}
\usepackage{float}
\usepackage[utf8]{inputenc}
\newcommand{\bfp}{\mathbf{p}}
\newcommand{\ncos}{N_{\cos\theta}}
\newcommand{\kbar}{\Bar{k}}
\newcommand{\omegabar}{\Bar{\omega}}
\newcommand{\tbar}{\Bar{t}}

\newcommand{\ipe}{{i_p}}
\newcommand{\itheta}{{i_\theta}}
\newcommand{\jpe}{{j_p}}
\newcommand{\jtheta}{{j_\theta}}

\title{Hydrodynamic and Non-hydrodynamic Excitations in Kinetic Theory -- A Numerical Analysis in Scalar Field Theory}
\author{Stephan Ochsenfeld,}
\emailAdd{s.ochsenfeld@uni-bielefeld.de}
\author{S\"{o}ren Schlichting}
\affiliation{Fakult\"{a}t f\"{u}r Physik, Universit\"{a}t Bielefeld,\\ Universit\"{a}tsstrasse 25, D-33615 Bielefeld, Germany}
\date{August 2023}

\begin{document}

\abstract{
Viscous hydrodynamics serves as a successful mesoscopic description of the Quark-Gluon Plasma produced in relativistic heavy-ion collisions. In order to investigate, how such an effective description emerges from the underlying microscopic dynamics we calculate the hydrodynamic and non-hydrodynamic modes of linear response in the sound channel from a first-principle calculation in kinetic theory. We do this with a new approach wherein we  discretize the collision kernel to  directly calculate eigenvalues and eigenmodes of the evolution operator. This allows us to study the Green's functions at any point in the complex frequency space. Our study focuses on scalar theory with quartic interaction and we find that the analytic structure of Green's functions in the complex plane is far more complicated than just poles or cuts which is a first step towards an equivalent study in QCD kinetic theory.
}

\maketitle

\clearpage

\section{Introduction} \label{sec:introduction}

Hydrodynamics play a central role in describing  the collective behavior of macroscopic systems in the real world. The premise of hydrodynamics is that it describes the late time and long wavelength limit of a system. The Quark-Gluon-Plasma (QGP) found in ultrarelativistic Heavy Ion Collisions (HIC) is very short lived and the average system size is small, yet data collected from heavy-ion collisons at RHIC and LHC suggests that the space-time dynamics can be well modelled by hydrodynamic theories~\cite{ALICE:2014dwt,ATLAS:2017hap,CMS:2017kcs,Nagle:2018nvi}. \\
Considering that the QGP is initially far from equilibrium and is subject to large gradients, naturally the question arises on what time and distance scales hydrodynamic theories can provide a realistic description of HICs~\cite{Baier:2006um,Teaney:2009qa,Song:2010mg,Snellings:2011sz,Schenke:2011zz,Heinz:2013th,Luzum:2013yya,Gale:2013da,Berges:2013fga,Bozek:2011if,Ambrus:2021fej,Ambrus:2022qya,Schenke:2022zkw}. Since in practice, hydrodynamic theories are based on expansions around local thermal equilibrium \cite{Denicol:2010xn,Kovtun:2012rj,Romatschke:2017ejr}, it has long been believed that proximity to equilibrium is a necessary criterion for the applicability of a fluid dynamic description. However, in recent years various studies have indicated that  -- at least for certain classes of microscopic systems -- viscous hydrodynamics can provide a rather accurate description even when the system is significantly out of equilibrium, featuring for example pressure anisotropies of order unity~\cite{Casalderrey-Solana:2017zyh,Blaizot:2019scw}. Hence it has become customary in the field of high-energy nuclear physics, to carefully distinguish equilibration from hydrodynamization, which merely refers to the applicability of viscous relativistic hydrodynamics, and we refer the interested reader to \cite{Casalderrey-Solana:2011dxg,Heller:2013oxa,Denicol:2014xca,Heller:2015dha,Kurkela:2015qoa,Romatschke:2016hle,Romatschke:2017vte,Berges:2020fwq} for recent perspectives on this issue. 
One field of study that emerged relatively recently to tackle the question of (in)applicability of hydrodynamics, consists of analyzing so called hydrodynamic and non-hydrodynamic modes \cite{Romatschke:2015gic,Grozdanov:2016fkt,Kurkela:2017xis,Moore:2018mma,Soloviev:2021lhs}. Hydrodynamic modes, which correspond to isolated singularities in the Fourier transformed evolution equations of hydrodynamic theories, are calculated by linearly perturbing evolution equations and calculating the system response. These modes fulfill the hydrodynamic limit $\lim_{k\rightarrow 0}\omega=0$, corresponding to conserved quantities. Non-hydrodynamic modes in turn do not fulfill this large wavelength limit and are present on any length scale, albeit varying in importance~\cite{Brewer:2022ifw,Ke:2022tqf,Du:2023bwi}. 
Although they are called non-hydrodynamic, they are also found in hydrodynamic theories such as Müller-Israel-Stewart \cite{Muller:1967zza,Israel:1979wp,Baier:2007ix,Romatschke:2009im,Denicol:2012cn}. So when these non-hydrodynamic modes are ever present, the question still stands when they dominate the dynamics of the system. The regime, where they are non-negligible, will help understand the boundaries of the applicability of hydrodynamics. \\
A common procedure to calculate the dynamics of the QGP are multi-stage models that use kinetic theories to describe the pre-equilibrium phase which then transition smoothly into the hydrodynamic regime \cite{Arnold:2002zm,Kurkela:2014tea,Blaizot:2014jna,Xu:2014ega,Scardina:2014gxa,Kurkela:2018wud,Kurkela:2018oqw,Carzon:2023zfp}. Kinetic theories are also sometimes used to directly derive new hydrodynamic theories, such as the DNMR hydrodynamic theory \cite{Muronga:2006zx,Betz:2009zz,Denicol:2010xn,Denicol:2012cn}.
Because of this ability to capture both non-equilibrium and non-hydrodynamic behavior kinetic theory is a fitting model to find non-hydrodynamic modes that are so natural to the system. Solving the Boltzmann equation analytically is not feasible for non-trivial kinetic theories, thus we have to solve the problem numerically. The problem then is to find the modes we are looking for. Analytically the modes are extracted as non-analytic structures of the Green's functions of the energy-momentum tensor, for example poles or cuts, but numerically it is hard to do the same. Going into frequency space becomes problematic numerically when the integral is not restricted to the real frequency axis. Therefore we need some other method to determine the complex structure of energy-momentum Green's functions. 

In this paper, we present a method to calculate eigenfunctions and eigenvalues of the Boltzmann equation with non-zero gradients $k$ in a discretized fashion by using moments of the distribution function. Based on the method described in Sec.\ref{seq:methodology}, we can discuss the location of singularities and the analytic structure of Green's function in the complex frequency plane. In Sec.\ref{seq:RTA} the procedure is tested within the Relaxation Time Approximation (RTA) and subsequently applied to scalar $\phi^4$ kinetic theory in Sec.\ref{seq:scalartheory}. The RTA can be compared to exact calculations and reproduce known results. The scalar calculations expand the known knowledge of zero gradient results into a finite $k$ regime. We find an analytic structure that goes beyond poles and cuts which was predicted by some previous works~\cite{Kurkela:2017xis,Moore:2018mma}.

\section{Methodology} \label{seq:methodology}

Kinetic theory is an effective mesoscopic theory, where the time evolution of the phase-space distribution $f(\bfp,\vec{x},t)$, where $\bfp=(p^0,\vec{p})=(p,\vec{p})$ is the four-momentum and $\mathbf{x}=(x^0,\Vec{x})=(t,\vec{x})$ are the space-time coordinates, is governed by a Boltzmann type equation. Within kinetic theory, it is comparatively straightforward to calculate the time evolution of the phase-space distribution, as well as the behavior of macroscopic quantities, such as e.g. retarded correlation functions of the energy-momentum tensor $G(t)$, which can be obtained from moments of the phase-space distribution. In particular, there are several numerical studies of kinetic theories, including scalar $\phi^4$-theory or even QCD kinetic theory \cite{Moore:2007ib,York:2008rr,Kurkela:2014tea,Scardina:2014gxa,AbraaoYork:2014hbk,Kurkela:2018wud,Moore:2018mma,Du:2020dvp,Mullins:2022fbx,Du:2023bwi}, which explore the real-time dynamics of the system. However, as mentioned before, the problem of analyzing the structure of response functions in the complex frequency space originates from the need for a Laplace transform 
$$
G(\omega)=\int_{0}^{\infty} dt~e^{i \omega t}~G(t)
$$
which becomes numerically ill behaved for complex frequencies with $\text{Im}(\omega)<0$, as the above integral does not converge beyond the first singularity in the lower half-plane. While for analytic solutions of the energy momentum Green's function it is still possible to calculate the Laplace transform directly, as has been done e.g. in RTA \cite{Romatschke:2015gic,Kurkela:2017xis}, this is evidently not possible with numerical data for $G(t)$, and thus a different approach is required.

\subsection{From Collision Integral to Matrix} 

In our method we will discretize the Boltzmann equation in order to calculate discrete spectra of eigenvalues and their eigenfunctions, which tell us exactly how the system can respond to perturbations. The general form of the Boltzmann equation is
\begin{align}
p^\mu \partial_\mu f(\bfp,\vec{x},t)=pC[f](\bfp,\Vec{x},t) \ ,
\end{align}
where $C[f]$ is the the collision operator, also called collision integral or collision term in the following, of the distribution function $f$. Throughout this work we use the mostly minus metric. We will study linear perturbations on top of an equilibrium background, which is defined by ultrarelativistic bosons with zero mass obeying the equilibrium distribution
\begin{align}
f_{\rm eq}(p)=n(p)=\frac{1}{e^{p/T}-1} \ .
\end{align}
Our perturbation is then of the form
\begin{align}
f(\bfp,\Vec{x},t)=n(p)+\delta f(\bfp,\vec{x},t).
\end{align}
The collision integral, in this study only for $2\leftrightarrow2$ scatterings, also has to be linearized around the equilibrium background, yielding
\begin{align}
C^{2\leftrightarrow 2}[\delta f](\bfp_1,t)=\frac{1}{2p_1}\frac{1}{2}\int \frac{d^3p_2d^3p_3d^3p_4}{(2\pi)^92p_22p_32p_4}(2\pi)^4\delta^4\left(\bfp_1+\bfp_2-\bfp_3-\bfp_4\right)|\mathcal{M}|^2\delta \mathcal{F} \ ,
\end{align} 
where $|M|^2$ is the scattering matrix element squared and $\delta \mathcal{F}$ is the linearized statistical factor
\begin{align}
\begin{aligned}
\delta \mathcal{F}=&-\delta f(\bfp_1,\vec{x},t)\left(n_2(1+n_3+n_4)-n_3n_4\right)-\delta f(\bfp_2,\Vec{x},t)\left(n_1(1+n_3+n_4)-n_3n_4\right)\\
&+\delta f(\bfp_3,\Vec{x},t)\left(n_4(1+n_1+n_2)-n_1n_2\right)+\delta f(\bfp_4,\Vec{x},t)\left(n_3(1+n_1+n_2)-n_1n_2\right) \ .
\end{aligned}
\end{align}
Note that we abbreviated the distribution functions in a way to show their dependence on a specific momentum, e.g. $n_1=n(p_1)$.
For RTA the collision term takes a much simpler form, which is discussed in the respective section. For the spatial part of the Boltzmann equation we switch into the Fourier space by transforming according to
\begin{align}
\delta f_k(\bfp,t)=\int d^3x~ e^{-i\vec{k}\cdot\Vec{x}}\delta f(\bfp,\Vec{x},t),
\end{align}
where gradients are replaced by the wave vector $\Vec{k}$
\begin{align}
\left(\partial_t +i\Vec{p}\cdot\Vec{k}\right)\delta f_k(\bfp,t)= C[\delta f_k](\bfp,t).
\end{align}
Fourier transformed quantities are marked with a $k$ in the subscript. In this work we will focus only on perturbations in the sound channel. For this we will initially perturb the system by a temperature perturbation, such that
\begin{align}
\delta f_k(\bfp,t=0)=\delta T_0\partial_T f_{\rm eq}=\frac{\delta T_0}{T}\frac{p}{T}e^{p/T}f_{\rm eq}^2 \ ,
\label{eq:initial_condition}
\end{align}
where $\delta T_0$ is the magnitude of our initial perturbation in $T$.
Since our background is isotropic and we will consider only scalar perturbations the perturbations will only depend on the absolute value of the wave vector $|\Vec{k}|$. Without loss of generality we orient the wave vector along the $z$-axis to receive the linearized Boltzmann equation
\begin{align}
\left(\partial_t+ik\cos\theta\right)\delta f_k(\bfp,t)= C[\delta f_k](\bfp,t)  ,
\label{eq:linearized_boltzmann}
\end{align}
where $\cos\theta=\frac{\Vec{k}\cdot\Vec{p}}{kp}$ is the longitudinal momentum angle and $k=|\Vec{k}|$. \\
The collision and gradient term will be discretized in order to solve it numerically. How we then extract and evaluate the eigenvalues depends on the diagonalization of said discretized collision integral. \footnote{In the following technical details we omit the gradient part of the Boltzmann equation just to illustrate how the discretization will work.} The discretized version of a linear operator is a finite dimensional matrix, which is easily decomposed into its eigenvectors. The eigenvectors of a matrix can be divided into right and left eigenvectors, which are usually not the same. Hermitian or symmetric matrices do not have this separation of eigenvectors. In the case of $k=0$ the evolution matrix would be symmetric, but the addition of gradient terms in the matrix makes it neither hermitian nor symmetric, meaning we have to respect the distinction of left and right eigenvectors. If $\Vec{a}_i$ is the right and $\Vec{b}_i$ the left eigenvector to the eigenvalues $\lambda_i$ of the matrix $C$ they satisfy
\begin{align}
C\Vec{a}_i=&\lambda_i\Vec{a}_i , \\
\Vec{b}_i^TC=&\lambda_i\Vec{b}_i^T.
\end{align}
The eigenvector can also be used to denote the diagonalization of $C$ via the modal matrix $P$
\begin{align}
C=PDP^{-1},
\end{align}
where $D=\text{diag}(\lambda_i)$ is the diagonal form of $C$. The modal matrix and its inverse are related to the eigenvectors via the following relations
\begin{align}
P_{ij}=(\Vec{a}_j)_i \ , \qquad P^{-1}_{ij}=(\Vec{b}_i)_j.
\end{align}
Since the discretized version of our distribution function will be a vector let us call it $\vec{f}$ here to illustrate the basic principle of the method. After the discretization process the Boltzmann equation will transform into a vector matrix equation
\begin{align}
\partial_t f=C[f] \ \rightarrow \ \partial_t \Vec{f}=C\Vec{f},
\end{align}
which is solved by $\Vec{f}(t)=e^{Ct}\Vec{f}(0)$. Using the diagonal form of $C$ and the representation of the modal matrix we receive 
\begin{align}
\Vec{f}(t)=Pe^{Dt}P^{-1}\Vec{f}(0)=\sum_i e^{\lambda_i t}\Vec{a}_i \left(\Vec{b}_i\cdot \Vec{f}(0)\right). \label{eq:eigenzerlegung}
\end{align}
This form gives the solution to the Boltzmann equation as a sum of contributions from each eigenvalue with some weight, that is dependent on the initial condition, which makes it incredibly easy to discuss the individual influence of each eigenvalue. 
From there it is also straight forward to go into frequency space  
\begin{align}
\vec{f}(\omega)=-\sum_i\frac{\vec{a}_i\left(\Vec{b}_i\cdot \vec{f}(0)\right)}{i\omega+\lambda_i}. \label{eq:laplacedecompose}
\end{align}
This form shows that each eigenvalue directly corresponds to a singularity in frequency space located at $\omega_i=-\text{Im}(\lambda_i)+i\text{Re}(\lambda_i)$, revealing the complex structure of the theory.

\subsection{Moments of the Distribution Function}
We just showed how an operator equation may be rewritten into a much simpler and easier analyzable form. We will now discuss how the distribution function and collision integral are transformed into a discrete space. \\
We start by the introduction of moments of the distribution function 
\begin{align}
N_i(t)=\int \frac{d^3p}{(2\pi)^3}\delta f_k(\bfp,t)w_i(\bfp) \ ,
\end{align}
where the $w_i(\bfp)$ are some weight functions. This way one can construct a vector $\Vec{N}$, where the components are the individual moments $N_i$. 
Our approach follows \cite{AbraaoYork:2014hbk,Soudi:2021aar} in the construction of their weight functions. Since we will only study sound mode excitations, it is sufficient to discretize the momentum $p$ and the polar angle $\cos\theta$. We do this with the use of so called wedge functions $w_i(x)$, which are defined as
\begin{align}
w_i(x)=\left\{\begin{matrix} \frac{x-x_{i-1}}{x_i-x_{i-1}}\ , & x_{i-1}<x<x_i \\ \frac{x_{i+1}-x}{x_{i+1}-x_i}\ , & x_i<x<x_{i+1} \\ 0 \ , & \text{else} & \end{matrix} \right.
\end{align}
where the $x_i$ are discrete grid points of the quantity that is discretized in the wedge function. The momentum grid contains momenta from $p_{min}=0$ to $p_{max}$ evenly spaced and $N_p$ in number. The $\cos\theta$ grid contains $N_{\cos\theta}$ points and goes from $\cos\theta_{min}=-1$ to $\cos\theta_{max}=1$. This means the grid has a total size of $N_{tot}=N_pN_{\cos\theta}$ entries. The wedge functions fulfill following relations
\begin{align}
\begin{aligned}
\sum_i w_i(x)&=\Theta(x_{max}-x)\Theta(x-x_{min}) \ , \\
\sum_i x_iw_i(x)&=x\Theta(x_{max}-x)\Theta(x-x_{min}) \ . \label{eq:wedgeconservation}
\end{aligned}
\end{align}
So the moments of the distribution function, also sometimes called wedge moments in the following, are then defined as
\begin{align}
N_i(t)=\int \frac{d^3p}{(2\pi)^3} w_{i_p}(p)w_{i_\theta}(\cos\theta)\delta f_k(\bfp,t),
\end{align}
where $i$ is a combined index in the form of $i=i_\theta+i_pN_{\cos\theta}$. Using the properties of the wedge function one can gather information about particle number $\delta n_k=\delta J^0_k$, energy $\delta e_k=\delta T^{00}$ and longitudinal momentum $\delta\pi_k=\delta T^{03}_k$ by simply summing over the wedge moments in the form of
\begin{align}
\delta n_k(t)=\sum_i N_i(t) \ , \quad \delta e_k(t)=\sum_i p_{i_p} N_i(t) \ , \quad \delta\pi_k(t)=\sum_i p_{i_p}\cos\theta_{i_\theta}N_i(t), \label{macroreconstruction}
\end{align}
where we used the particle four current $\delta J_k^\mu$ and the energy momentum tensor $\delta T^{\mu\nu}$, calculated in kinetic theory by taking moments of the distribution function
\begin{align}
\delta J_k^\mu(t)&=\int\frac{d^3p}{(2\pi^3)}\frac{p^\mu}{p}\delta f_k(\bfp,t) \ , \\
\delta T^{\mu\nu}_k(t)&=\int\frac{d^3p}{(2\pi)^3}\frac{p^\mu p^\nu}{p}\delta f_k(\bfp,t) \ . \label{eq:energy_momentum_tensor}
\end{align}
We still need a way to calculate the matrix described in our method. We do this in the same moment approach and take moments of the collision integral 
\begin{align}
C_i(t)=\int \frac{d^3p}{(2\pi)^3}w_{i_p}(p)w_{i_\theta}(\cos\theta)C[\delta f_k](\bfp,t) \ .
\end{align}
In order to construct a matrix from this we have to invert the wedge moments back into the form of a distribution function. We do this by approximation of the original moment integral. When we take $\frac{\delta f_k(\bfp,t)}{f_{\rm eq}e^{p/T}}$ to be constant between nodes we can rewrite the Integral as 
\begin{align}
N_i(t)=\int \frac{d^3p}{(2\pi)^3}w_{i_p}(p)w_{i_\theta}(\cos\theta)\delta f_k(\bfp,t) \frac{f_{\rm eq}(p)e^{p/T}}{f_{\rm eq}(p)e^{p/T}}=\frac{\delta f_k(\bfp,t)}{f_{\rm eq}(p)e^{p/T}} A_i,
\label{eq:momentapproximation}
\end{align}
where we introduced a new area function $A_i$ that is calculated as
\begin{align}
A_i=\int \frac{d^3p}{(2\pi)^3}w_\ipe(p)w_\itheta(\cos\theta)f_{eq}(p)e^{p/T} \ .
\end{align}
Eq.(\ref{eq:momentapproximation}) only holds for the surrounding of the $i-$th node point, thus in order to recover the whole distribution function we have to sum over all wedges
\begin{align}
\delta f_k(\bfp,t)=&\sum_i w_\ipe(p)w_\itheta(\cos\theta)N_i(t)\frac{f_{\rm eq}(p)e^{p/T}}{A_i}
=\sum_i K_i(\bfp)N_i(t) \ ,
\end{align}
where we introduced the "Cardinal Function" $K_i(\bfp)$
\begin{align}
K_i(\bfp)=w_\ipe(p)w_\itheta(\cos\theta)\frac{f_{\rm eq}(p)e^{p/T}}{A_i} \ .
\end{align}
With the introduction of this Cardinal Function the Collision integral moments become linear functions of the distribution moments and thus we can rewrite the equation as a matrix vector multiplication
\begin{align}
C_i(t)=\sum_j C_{ij}N_j(t)=\left(C\vec{N}(t)\right)_i \ .
\end{align}
The matrix entries are then easily calculated with the functional derivative \begin{align}
C_{ij}=\frac{\delta C_i(t)}{\delta N_j(t)} \ .
\end{align}
With these preparations we can rewrite the Boltzmann equation for $k=0$ as a matrix equation
\begin{align}
\partial_t \vec{N}(t)=C\vec{N}(t) \ ,
\end{align}
which is the desired form.\\
In addition to the collision integral the Boltzmann equation contains parts with nonzero $k$. These parts also have to be translated into the moment space. Like the collision integral they are linear in the distribution function and thus we can calculate a matrix in a similar fashion. The moments of the gradient contribution are given as
\begin{align}
M_i=-ik\int \frac{d^3p}{(2\pi)^3}\cos\theta w_\ipe(p)w_\itheta(\cos\theta)\delta f_k(\bfp,t) \ .
\end{align}
from which we calculate the matrix $M$ as
\begin{align}
M_{ij}=\frac{\delta M_i(t)}{\delta N_j(t)} \ .
\end{align}
Note that the $k$ dependent part is independent of $p$, thus one only needs to construct angular wedge moments. Then the matrix elements can be written as
\begin{align}
M_{ij}=M^{cos\theta}_{i_{\theta}j_{\theta}}\delta_{i_{p}j_{p}}
\end{align}
where $M^{cos\theta}_{ij}$ are now the angular wedge moments of the gradient part. The Kronecker Delta part is there to assure that no momenta are mixing. The angular wedge moments are defined as
\begin{align}
M^{cos\theta}_{ij}=-\frac{ik}{A^{\cos\theta}_j}\int_{-1}^1 d\cos\theta~w_{i_{\theta}}(\cos\theta)w_{j_{\theta}}(\cos\theta)\cos\theta \ ,
\end{align}
which can be easily calculated by hand. The $A^{\cos\theta}_j$ are area functions for the angular wedge moments just like the recent $A_j$ but are much simpler

\begin{align}
A^{\cos\theta}_j=\left\{\begin{matrix} \frac{1}{2}\Delta\cos\theta\ , & j=1 , j=N_{\cos\theta} , \\ \Delta\cos\theta\ , & \text{else} & \end{matrix} \right. \ .
\end{align}
$\Delta\cos\theta=\cos\theta_{i+1}-\cos\theta_i$ is the distance between each angular grid point, which is constant throughout because we choose an evenly spaced grid. This allows for a fast calculation of the gradient contribution of the equation, since only $k$ has to be multiplied to known matrix elements. In contrast the calculation of the collision integral matrix is done with a Monte Carlo scheme, where we update all matrix elements simultaneously with each sampling, such that particle number and energy momentum conservation are ensured with the help of the wedge function properties in Eq.(\ref{eq:wedgeconservation}). \\ 
Both collision integral and gradient term are combined as matrices in the ordinary differential equation
\begin{align}
\partial_t \Vec{N}(t)=(C+M)\Vec{N}(t),
\end{align}
with the solution
\begin{align}
\Vec{N}(t)=e^{(C+M)t}\Vec{N}(0) \label{eq:ode_sol}.
\end{align}
Calculating the collision integral matrix in high precision takes some time but can be saved for reuse in the same discretization because for different $k$ only the gradient contribution changes. This also saves a lot of computation time where the remaining computation time is due to the numeric calculation of the eigenspace.

\subsection{Observables and Complex Frequencies}
We now have the means to completely discretize the Boltzmann equation including gradient contributions. In the following we discuss the construction of observables from moments and the possibility to go into the complex frequency plane. \\
Observables that are linear moments of the distribution function $\delta f$, like energy, can be easily retrieved (see Eq.(\ref{macroreconstruction})) from wedge moments via a sum of moments in addition to some weight. We can write a representation of the energy in our moment space as a vector 
\begin{align}
\vec{O}_e=\begin{pmatrix} p_{1_p} \\ p_{2_p} \\ \vdots \\ p_{N_{tot_p}}
\end{pmatrix} \ ,
\end{align}
because we get the energy by forming the scalar product 
\begin{align}
\delta e_k(t)= \vec{O}_e\cdot \vec{N}(t) \ .
\end{align}
The Green's function is the time evolution of an observable as a response to an initial perturbation of some sort. This initial condition is already encoded in the time evolution of $\Vec{N}(t)$ as $\Vec{N}(0)$(see Eq.(\ref{eq:ode_sol})). Thus we then define the Green's function as the scalar product of the corresponding observable vector with the distribution vector
\begin{align}
G(t)=\vec{O}\cdot \vec{N}(t) \ .
\label{eq:observable}
\end{align}
The full form of the Green's function can then be calculated using the solution to the ODE in Eq.(\ref{eq:eigenzerlegung}) as
\begin{align}
G(t)=\sum_i e^{\lambda_i t}\left(\vec{O}\cdot\vec{a}_i\right)\left(\vec{b}_i\cdot\vec{I}\right) \ . \label{eq:real_green_decomposed}
\end{align}
Each eigenvalue is behaving as a complex exponential function with a certain weight, we call this the contribution or residue to the Green's function. Let us define the contribution of an eigenvalue as $\mu_i$
\begin{align}
\mu_i=\left(\vec{O}\cdot\vec{a}_i\right)\left(\vec{b}_i\cdot\vec{I}\right) \ .
\label{eq:residue}
\end{align}
The Laplace transform of the Green's function is then
\begin{align}
G(\omega)=-\sum_i \frac{\mu_i}{i\omega + \lambda_i} \ .\label{eq:freq_green_decomposed}
\end{align}
With this we have a direct way to study the Green's function as a function of time or in the complex frequency plane, where each eigenvalue is an individual singularity.

\subsection{Scaling Behavior}
Since we are studying the theory with an approach to hydrodynamics in mind it makes sense to use hydrodynamic scaling properties. In first order viscous hydrodynamic theory 
\begin{align}
G(t)=\cos(c_s kt)e^{-\frac{2}{3}\frac{\eta}{sT}k^2t}
\end{align}
one can rescale the wavenumber $k$ and time $t$ by the viscosity to receive a universal description independent of viscosity \cite{Du:2023bwi}. Thus we define the rescaled wavenumber, time and frequency as
\begin{align}
    \Bar{k}=k\frac{\eta}{sT} \ , \quad \Bar{t}=t\frac{sT}{\eta} \ , \quad \Bar{\omega}=\omega\frac{\eta}{sT}.
\label{eq:scaling_variables}
\end{align}
Notice that the combined $\Bar{k}\Bar{t}$ is still equal to $kt$ and $\Bar{\omega}$ is defined according to $\Bar{t}$, being the Fourier counterpart of $\Bar{t}$. Except for the scaling variables, all dimensionful variables, like $p_{max}$, are expressed in units of the only dimensionful scale, the background temperature $T$.

\section{Benchmark with RTA} \label{seq:RTA}
Before we apply the method to scalar theory we should put it to the test within a well studied kinetic theory, the RTA~ \cite{Kurkela:2017xis,Romatschke:2015gic}. In RTA all excitations of different momenta decay equally with a rate $\tau_R$, the relaxation time, correlating to a branch cut in the retarded Green's function. The RTA Boltzmann Equation is of the form
\begin{align}
p^\mu\partial_\mu f(\bfp,\vec{x},t)=\frac{p^\mu u_\mu}{\tau_R}\left(f_{\rm eq}(p^\mu u_\mu)-f(\bfp,\vec{x},t)\right), \label{eq:boltzmann_rta}
\end{align}
where the local rest frame velocity $u^\mu$ is determined by Landau matching as eigenvector of the background energy momentum tensor
\begin{align}
T^{\mu \nu}_{eq} u_\nu = e_{eq}u^\mu.
\end{align}

\subsection{Analytical Results}
To study the linear response to a perturbation we first use the linearized version of Eq.(\ref{eq:boltzmann_rta})
\begin{align}
    p^\mu\partial_\mu\delta f(\bfp,\vec{x},t)=\frac{p}{\tau_R}\left(\delta f_{\rm eq}(\bfp,t)-\delta f(\bfp,\vec{x},t)\right) \ .
\end{align}
For positions we switch into Fourier space, without loss of generality orienting the wave vector $\vec{k}$ only along the $z$-axis. For time we do a Laplace transform instead of Fourier transform since we have strictly positive times and want to include initial conditions. This results in
\begin{align}
-i\omega\delta f_k(\bfp,\omega) - \delta f_k(\bfp,t=0) +ik\cos\theta \delta f_k(\bfp,\omega)+\frac{1}{\tau_R}\delta f_k(\bfp,\omega)=\frac{1}{\tau_R}\delta f_{k,\rm eq}(\bfp,t) \ , \label{eq:perturbed_boltzmann_rta}
\end{align}
where $\delta f_k(\bfp,t=0)$ is the initial condition of the distribution function. The perturbed equilibrium distribution is given by gradients of temperature and velocity emerging from a change in $\delta f$ as
\begin{align}
    \delta f_{k,\rm eq}(\bfp,t)=\frac{p}{T}f_{\rm eq}^2\left(\frac{\delta T_k}{T}-\delta u^\mu_k v_\mu\right)e^{p/T} \ .
\end{align}
With this we can get the solution to Eq.(\ref{eq:perturbed_boltzmann_rta}) 
\begin{align}
\delta f_k(\bfp,\omega)=\frac{\delta f_{k,\rm eq}(\bfp,\omega)+\delta f_k(\bfp,t=0)\tau_R}{1+ik\tau_R\cos\theta-i\omega\tau_R} \ .
\end{align}
Via Landau matching we can relate the perturbed temperature and velocity to $\delta T^{\mu\nu}$ as
\begin{align}
\frac{\delta T_k}{T}&=\frac{1}{4}\frac{\delta e_k}{e}=\frac{1}{4e}\int \frac{d^3p}{(2\pi)^3}p \delta f_k \ ,
\\
\delta u^\mu_k &= \frac{\delta T^{0\mu}_k}{e+P}=\frac{3}{4e}\int \frac{d^3p}{(2\pi)^3}p^\mu \delta f_k \ .
\end{align}
Since our $\Vec{k}$ only lies in $z$-direction the only relevant components of the Energy-Momentum Tensor are $\delta T^{00}_k=\delta e_k$ and $\delta T^{03}_k$. The energy and velocity perturbation are related to the Green's functions we want to find by
\begin{align*}
G_{00,k}^{00}(\omega)=\frac{\delta e_k}{\delta e_0} \ , \quad G_{00,k}^{03}(\omega)=\frac{\delta T^{03}_k}{\delta e_0}.
\end{align*}
From the above equations one receives coupled equations for $\delta e_k$ and $\delta T^{03}_k$. The details of the solution for those are found in the Appendix \ref{appendix:rta_analytic}. When we define $L=\log(\frac{1-i\tau_R(k+\omega)}{1+i\tau_R(k-\omega)})$ the perturbed energy and longitudinal momentum are
\begin{align}
G_{00,k}^{00}(\omega)=\frac{-(6ik\tau_R+(3+k^2\tau_R^2-3i\omega\tau_R)L)}{(k^2\tau_RL+2ik^3\tau_R^2-6k\tau_R\omega+3\omega(i+\omega\tau_R)L)} \ , \\
G_{00,k}^{03}(\omega)=\frac{ik\tau_R(-2k\tau_R+(i+\omega\tau_R)L)}{(k^2\tau_RL+2ik^3\tau_R^2-6k\tau_R\omega+3\omega(i+\omega\tau_R)L)} \ .
\end{align}
Evidently, the above Green's functions feature a logarithmic branch cut extending between the branch points $\omega=-k-i/\tau_R$ and $\omega=k-i/\tau_R$ in the complex frequency plane. 
By expanding the inverse of the Green's function $G_{00,k}^{03}$ to second order in $\omega \tau_R$ one also finds a pair of hydrodynamic poles 
\begin{align}
\omega = \pm\frac{1}{\sqrt{3}}k-\frac{2}{15}ik^2\tau_R + \mathcal{O}(k^3)
\end{align}
for small frequencies and gradients, while for large gradients they disappear behind the cut as discussed in detail in \cite{Romatschke:2015gic}.

\subsection{Results from Numerical Method}
\begin{figure}[ht]
\centering
\includegraphics[width=\textwidth]{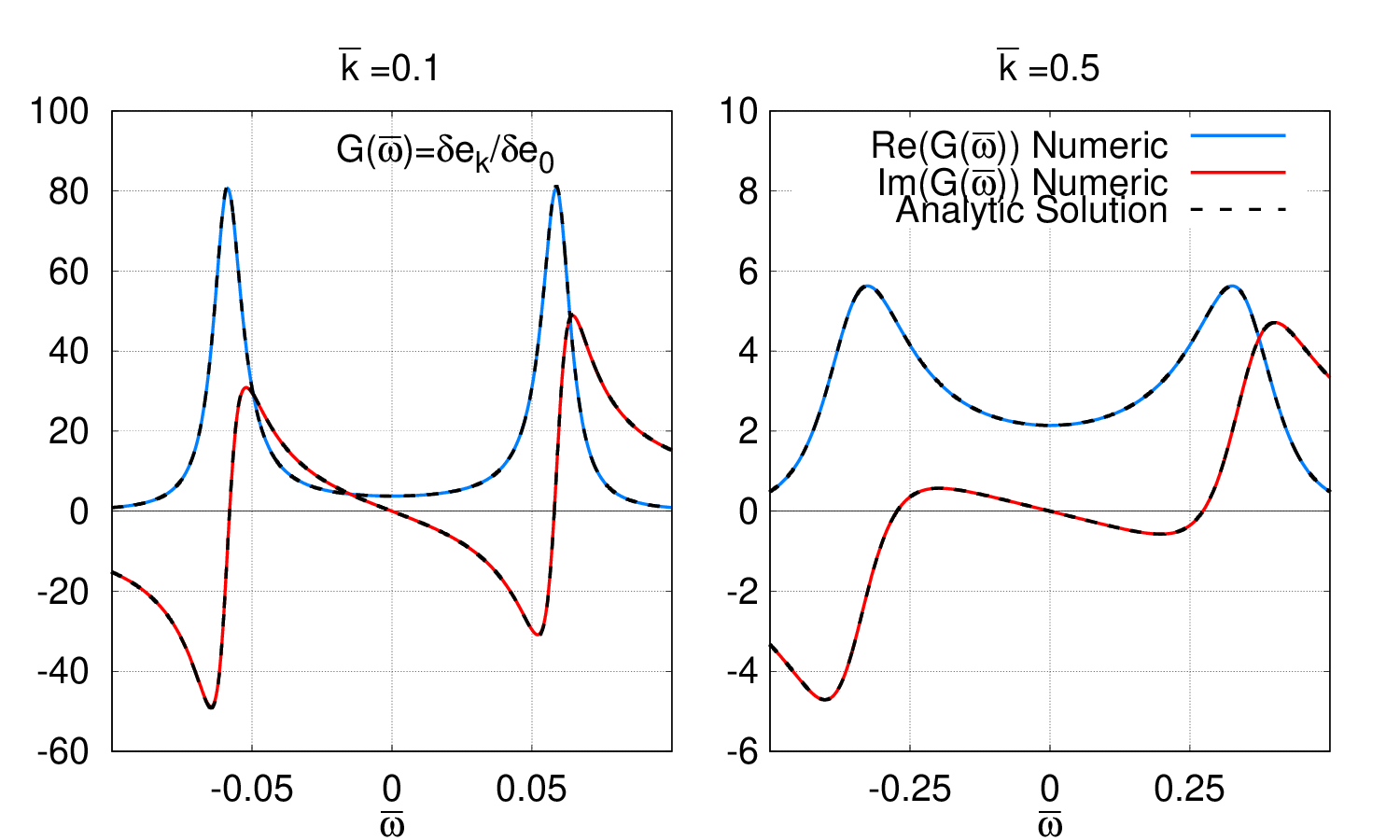}
\caption{Energy Green's function $G(\omegabar)$ in RTA as a function of real frequency $\omegabar$ for two different gradients $\kbar$. Analytic solution to the Boltzmann equation in black and the solution with our numerical method in red and blue.}
\label{fig:rta_comparison}
\end{figure}

Now that we have obtained the analytic solution, we can use it to benchmark our numerical method. Within the numerical method, we write the RTA Boltzmann equation as
\begin{align}
\partial_t \delta f_k(\bfp,t) + ik\cos\theta~ \delta f_k(\bfp,t)=\frac{1}{\tau_R} e^{p/T} f_{\rm eq}^2(p)\left(\frac{p}{T}\frac{\delta T_k}{T}-\frac{\delta u^\mu_k p_\mu}{T}\right)-\frac{1}{\tau_R}\delta f_k (\bfp,t)\ , \label{eq:perturbed_rta_timeevo}
\end{align}
which is then expanded in the moment approach as discussed in detail in Appendix \ref{appendix:rta_numerics}. 
We can now compare the results from our numeric method with the known analytic results in RTA. In Fig.\ref{fig:rta_comparison} we present the real and imaginary part of the energy Green's function plotted as a function of real frequency. The colored lines represent the numeric results from our new approach and the black dotted lines are the prior calculated analytic form. The numeric approach fully reproduces the analytic Green's function over the whole range. As increasing the gradient does not change this fact, we conclude that the new approach is a suitable method to calculate kinetic theory response functions.

Of course our main goal is to analyze the analytic structure in the complex frequency plane. Hence we still need to check if the new approach reproduces the well studied RTA structure, namely a cut and two poles, for complex frequencies.
\begin{figure}[ht]
    \centering
    \includegraphics[width=\textwidth]{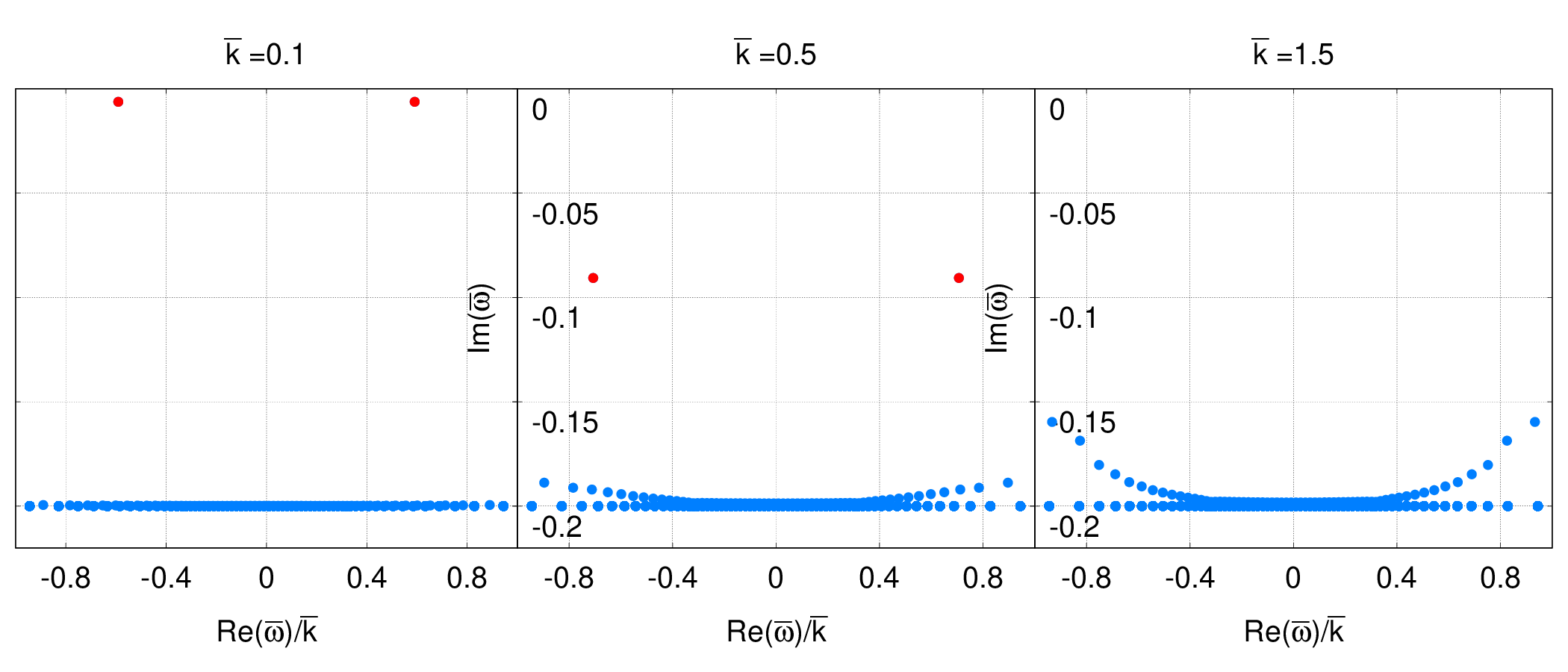}
    \caption{Eigenvalues of the RTA evolution operator for finite gradients $\kbar$ in the complex frequency plane. Hydrodynamic modes are highlighted in red.}
    \label{fig:rta_eigenvalues}
\end{figure}
In Fig.\ref{fig:rta_eigenvalues} we present each eigenvalue in the complex frequency plane for three gradients $\kbar$. We see clearly two isolated modes on top which correspond to the two well known hydrodynamic poles. The eigenvalues below form a line, which represents a branch cut in discretized fashion. The cut is located at $\bar{\omega}=-0.2i$ which coincides with the expected RTA cut located at $\omega=-i/\tau_R$ as the viscosity in RTA is given by $\eta/s=\frac{1}{5}\tau_R T$ \cite{Rocha:2022fqz,York:2008rr}. When $\kbar$ increases the hydrodynamic poles wanders towards the cut until they further disappear behind the cut. This behavior is also known from analytical results of RTA \cite{Romatschke:2015gic}. The deviations from a straight line occurring at higher $\kbar$ stems from discretization effects, which are discussed in the next chapter. Thus we conclude that the new approach shows the analytic structure clearly in the complex plane, which accurately describes the analytic knowledge.

\subsection{Effects of Discretization}

\begin{figure}[t]
\centering
\includegraphics[width=\textwidth]{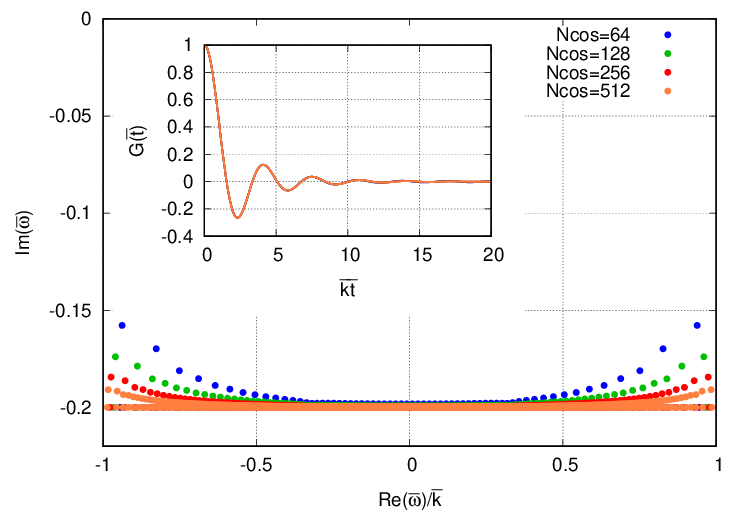}
\caption{Eigenvalues of the RTA evolution operator for various discretizations of $\ncos$ in the complex frequency plane. The inset plot shows the real time Green's function $G(\tbar)$ as a function of $\kbar\tbar$ for the same discretizations.}
\label{fig:rta_discretization}
\end{figure}
As we saw in the previous chapter the eigenvalues of the RTA should be two hydrodynamic poles and one cut located at $-i/\tau_R$. Additionally to the cut there are some aberrations at the edge of the cut that lean upwards in imaginary axis direction. To check if the deviations from the expected behavior are under control we analyze the effects of the discretization in the following. The increase of $N_p$ converges very quickly into a final solution with $N_p=64$ both in the time evolved Green's function $G(t)$ and the eigenvalue picture in the complex frequency plane. Thus we will only discuss the influence of $\ncos$ by varying $\ncos$ and holding $N_p$ constant at $N_p=16$. The results are presented in Fig.\ref{fig:rta_discretization}, where the eigenvalues for various discretizations are plotted in the complex frequency plane for $\Bar{k}=1$. We can directly see that by increasing $\ncos$ we push the deviation of the eigenvalues at the edges of the cut down towards the cut, telling us that these deviations from the expected cut are purely discretization effects. 
We also have to look at the effects of $\ncos$ on the time dependent Green's function $G(t)$. The results of this are found in the inset plot of Fig.\ref{fig:rta_discretization}, where the same coloring as the eigenvalue plot is used to display different discretizations. For $\ncos=64$ the time evolution already reached its limit because further increase $\ncos$ does not change the function or at least only minimally for very late times. 
We conclude that discretization effects can be erased in the time evolved Green's function entirely by using a feasible discretization. The eigenvalue picture however requires a high $\ncos$ to come close to the analytic cut expected in RTA, which is not numerically feasible under the consideration of also using $N_p=64$. The effects of the discretization are rather easy to discern from the physics here, thus we use a discretization of $N_p \times\ncos=64\times64$ and keep in mind that the slight arcs of the cut are artificial.

\section{Scalar Theory} \label{seq:scalartheory}
Previously we have shown that our method is suitable to analyze kinetic theories in both time and frequency domain. Now we will apply the method to a kinetic theory where the behavior for non-vanishing gradients ($\Bar{k}\neq 0$) has not been studied yet, the scalar field theory for quartic interaction, also known as $\phi^4$ theory.
Studies for $\Bar{k}=0$ have shown a complex structure in the form of a cut on the imaginary axis, meaning that except for the conserved quantities all excitations of the system decay on different time scales \cite{Moore:2018mma}. With the addition of gradients one expects the emergence of additional complex structures, including two isolated poles, which coincide with hydrodynamic poles for very small gradients $\bar{k}\ll 1$. These hydrodynamic poles obey a dispersion relation, which for example in second order hydrodynamic theory \cite{Baier:2007ix} is given by
\begin{align}
\omegabar=\pm c_s \bar{k} -i\frac{2}{3}\bar{k}^2 \pm \frac{2}{3c_s}\left(c_s^2\bar{\tau}_{\pi}-\frac{1}{3}\right)\bar{k}^3 . \label{eq:brsss}
\end{align}
where $c_s^2=1/3$ is the speed of sound and $\bar{\tau}_{\pi}\approx6.1$ is a second order transport coefficient \cite{York:2008rr}. 
The Lagrangian of scalar $\phi^4$ theory is defined by the real scalar field $\phi$ and the coupling strength $\lambda$
\begin{align}
\mathcal{L}[\phi,\partial_\mu \phi]=\frac{1}{2}\partial_\mu\phi\partial^\mu\phi-\frac{m^2}{2}\phi^2-\frac{\lambda}{24}\phi^4 \ .
\end{align}
s.t. following \cite{Moore:2018mma} the collision integral in kinetic theory then takes the form
\begin{align}
C[f](\bfp_1,t)&=-\frac{1}{2p_1}\frac{1}{2}\int \frac{d^3p_2d^3p_3d^3p_4}{(2\pi)^92p_22p_32p_4}(2\pi)^4\delta^4\left(\bfp_1+\bfp_2-\bfp_3-\bfp_4\right)\lambda^2 \\
&\times \left( f(\bfp_1)f(\bfp_2)\left[1+f(\bfp_3)\right]\left[1+f(\bfp_4)\right] - f(\bfp_3)f(\bfp_4)\left[1+f(\bfp_1)\right]\left[1+f(\bfp_2)\right] \right) \nonumber \ .
\end{align}
As explained before we want to use scaling variables in order to describe the system universally and independent on viscosity. The viscosity in scalar $\phi^4$ theory has been calculated numerically multiple times \cite{Jeon:1995zm,Moore:2007ib,Moore:2018mma} and for this work we choose the result from \cite{Jeon:1995zm}, which gives the viscosity in terms of temperature $T$ and coupling $\lambda$ as
\begin{align}
\eta \approx 3040\frac{T^3}{\lambda^2}.
\end{align}
To write this in a usable manner as viscosity $\eta$ over entropy density $s$ we use the thermodynamic relations of a massless gas of ultrarelativistic bosons \cite{Kovtun:2012rj} and receive
\begin{align}
\frac{\eta}{s}=\frac{45}{2\pi^2}\frac{\eta}{T^3}\approx \frac{6930}{\lambda^2}.
\end{align}
In this form it is also straightforward to see that, by use of the scaling variables defined in Eq.(\ref{eq:scaling_variables}), the coupling strength $\lambda$ can be completely scaled out of the Boltzmann equation Eq.(\ref{eq:linearized_boltzmann}), such that the results presented in the forthcoming sections are independent of the coupling strength.\\
We emphasize at this point, that the following results are obtained with an effective kinetic description of the scalar $\phi^4$ quantum field theory. By using an effective kinetic description, which can be derived from a combined weak-coupling and gradient expansion in quantum field theory~(see e.g. \cite{Berges:2005md}), we impose at least some restrictions to the possible eigenvalue picture in the complex frequency plane, such that for example the real part of the complex eigenvalues is restricted to the range ${-k<\rm Re}(\omega)<k$ \cite{Kurkela:2017xis}. Hence it is possible that the analytic structure of the real quantum field theory will differ from our kinetic theory results, as microscopic information is lost upon construction of the kinetic description~\cite{Weinstock:2005jw}.

\subsection{Spectrum for k=0}
Before we address the analytic structure of the Green's functions at finite wave-number,
the first result we should reproduce concerns the analytic structure of the Green's function in the absence of gradients ($\Bar{k}=0$). As already shown in \cite{Moore:2018mma} the Green's function for vanishing gradients exhibits a branch cut located on the imaginary axis. 
\begin{figure}[ht]
\centering
\includegraphics[width=0.9\textwidth]{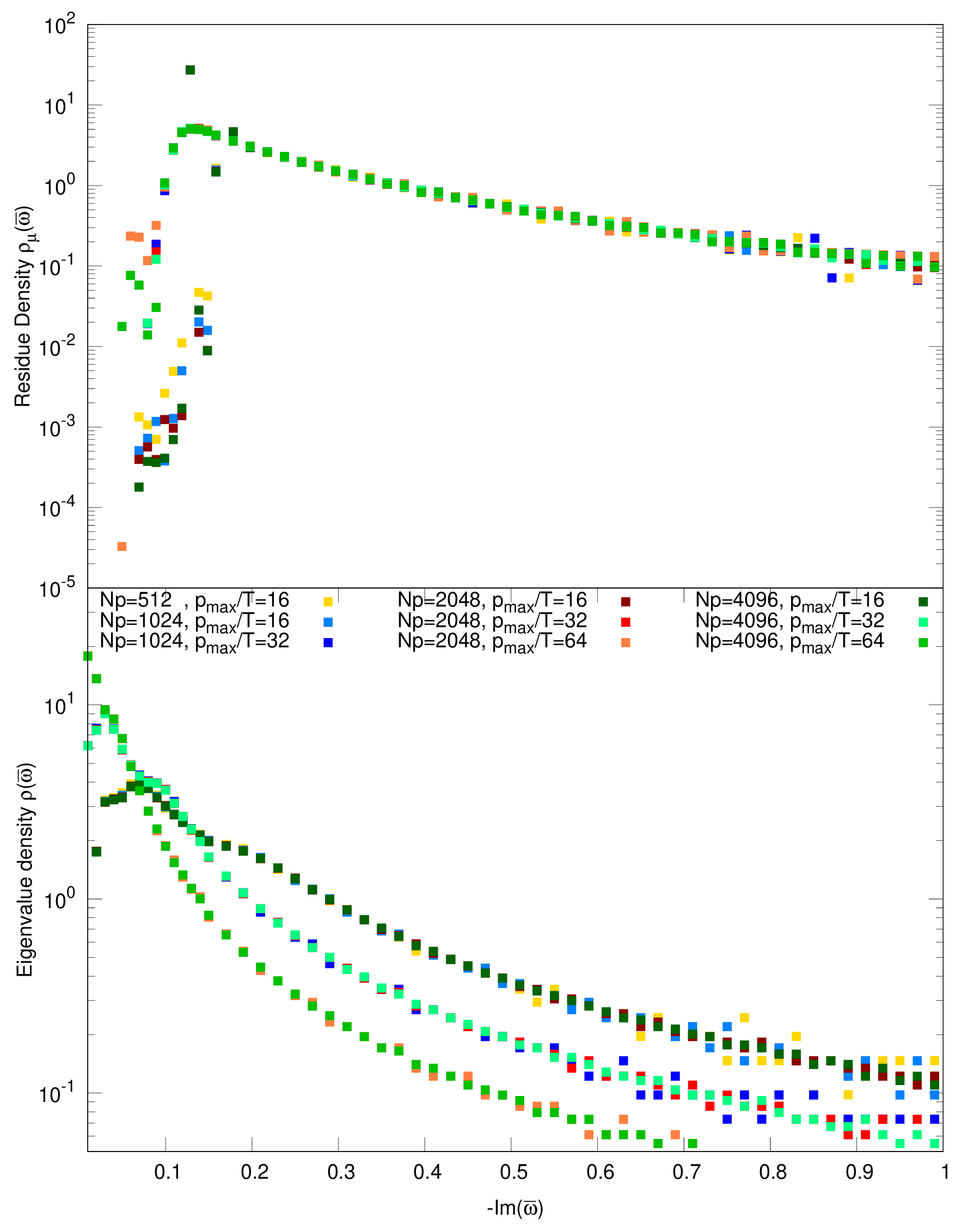}
\caption{Top: The residue density $\rho_\mu(\omegabar)$ for non-conserved observable in scalar $\phi^4$ theory for various discretizations as a function of imaginary frequency Im$(\omegabar)$. Bottom: The eigenvalue density $\rho(\omegabar)$ for the same discretizations as a function imaginary frequency Im$(\omegabar)$.}
\label{fig:scalar_zerok}
\end{figure}
In order to analyze this behavior, we investigate the spectrum of eigenvalues of the evolution operator, which in the absence of gradients only consists of the collision operator. Our results show three distinct zero modes which correspond to the conservation of energy, momentum and particle number in the sound channel. Other than that the spectrum only contains eigenvalues which are located on the imaginary frequency axis. In order to analyze whether this spectrum is discrete or continuous, one has to investigate how the eigenvalue density and their respective contribution to physical observables behaves in the continuum limit. \\
As objects to study these properties with we choose the eigenvalue density $\rho(\Bar{\omega})$  and residue density $\rho_\mu(\Bar{\omega})$. Since all eigenvalues have no real frequency $\text{Re}(\omegabar)$ the densities are calculated with respect to the imaginary frequency $\text{Im}(\omegabar)$. The eigenvalue density is calculated by counting all eigenvalues $\lambda_i$ in a frequency bin of size $\Delta \omegabar$ with the help of the step function $\Theta_{j}(\bar{\omega}_i,\omegabar)$ which is defined as
\begin{align}
\Theta_{j}(\bar{\omega}_i,\omegabar)=\Theta\big(\Delta\omegabar-2|\text{Im}(\omegabar)_i-(j+1/2)\Delta\omegabar|\big)\Theta\big(\Delta\omegabar-2|\text{Im}(\omegabar)-(j+1/2)\Delta\omegabar|\big),
\end{align}
where $\omegabar_i$ is the frequency of each eigenvalue, as seen in Eq.(\ref{eq:laplacedecompose}). Normalizing this by the total number of eigenvalues, ignoring the three zero modes, we receive the eigenvalue density
\begin{align}
\rho(\omegabar)=\sum_{i=1}^{N_{tot}-3}\sum_{j=0}^{\infty} \frac{\Theta_{j}(\bar{\omega}_i,\omegabar)}{(N_{tot}-3)\Delta\omegabar}.
\end{align}
In order to study a non-trivial residue density a suitable observable has to be chosen. All conserved charges have vanishing residue for $\omegabar\neq 0$, hence a non conserved charge has to be used. As observable we consequently choose the momentum squared
\begin{align}
\Vec{O}_{p^2}=\begin{pmatrix}
p_{1_p}^2 \\ p_{2_p}^2 \\ \vdots \\ p_{Ntot_p}^2
\end{pmatrix}.
\end{align}
The residue $\mu_i$ of this observable per eigenvalue is given by Eq.(\ref{eq:residue}). We have to sum this, analogously to the eigenvalue density, in a frequency bin $\Delta\omegabar$ and normalize it. Here we don't normalize by the total number of eigenvalues but the total residue of eigenvalues with $\omegabar\neq0$, which is subject to small variations due to the change in discretization\footnote{These variations are negligible and don't change the physics but rather disturb the purpose of this analysis.}. If we define this total non-conserved residue as
\begin{align}
\mu=\sum_{i=0}^{N_{tot}-3}\mu_i,
\end{align}
we can write the residue density of the non-conserved observable as
\begin{align}
\rho_\mu(\omegabar)=\sum_{i=1}^{N_{tot}-3}\sum_{j=0}^{\infty}\frac{\mu_i\Theta_{j}(\omegabar_i,\omegabar)}{\mu\Delta\omegabar}, \label{eq:residue_density}
\end{align}
where we also ignore the three zero modes in the sum. Since for our choice of initial condition (c.f Eq.(\ref{eq:initial_condition})), the angular structure is trivial for $\bar{k}=0$, the discretization analysis is carried out by varying the number of momentum points $N_p$, which refines the discretization and the momentum cut-off $p_{max}$, which can be used to explore the addition of high momentum modes. \\
We illustrate the results in Fig.\ref{fig:scalar_zerok} by plotting the eigenvalue density $\rho(\omegabar)$ in the bottom panel and the residue density $\rho_\mu(\omega)$ in the top panel. Each discretization is presented by a different color. The first thing to notice in the bottom panel is that the increase in $p_{max}$ corresponds to an addition of eigenvalues in the low frequency regime close to the origin, such that the number of eigenvalues decreases in higher frequency regions compared to discretizations with the same $N_p$. Vice versa when we fix $p_{max}$ and increase the number of momenta $N_p$ the density seems to approach a smooth continuum limit for each value of the momentum cut-off.
Since the variation in $p_{max}$ induces strong changes in the eigenvalue density we have to check if this influences the physical properties of the system. For this see the top panel of Fig.\ref{fig:scalar_zerok}. Here we also see that additional low frequency eigenvalues get added when we increase $p_{max}$ but all of them have exponentially small residue, thus making them negligible for the linear response of the system. The increase in $N_p$ has the same influence on the residue density as on the eigenvalue density. By simultaneously increasing $p_{max}$ and $N_p$, the residue density seems to approach a genuine continuum limit with better and better discretization. 
We conclude that in the continuum limit $N_p,p_{max}\rightarrow \infty$ the eigenvalues are continuously distributed along the negative imaginary frequency axis, representing a branch cut, which is in line with the prior works of Moore \cite{Moore:2018mma}. While the branch cut terminates only at the real frequency axis, the contribution of points close to the real axis to physical observables is exponentially suppressed, as the dominant contribution originates from points with $\text{Im}(\bar{\omega})\sim 0.1$.

\subsection{Effects of discretization}
\begin{figure}[ht]
\centering
\includegraphics[width=0.49\textwidth]{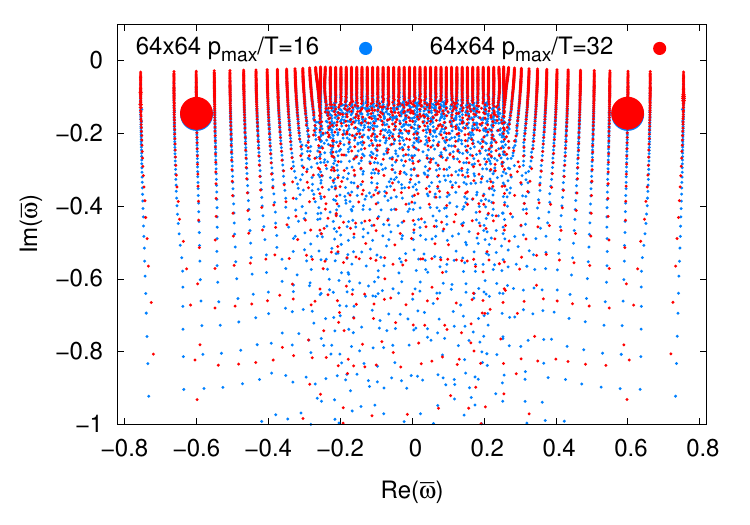}
\includegraphics[width=0.49\textwidth]{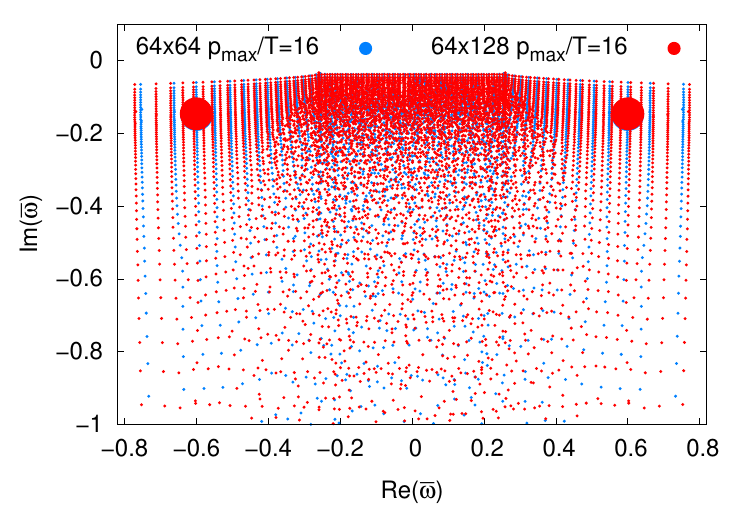}
\includegraphics[width=0.49\textwidth]{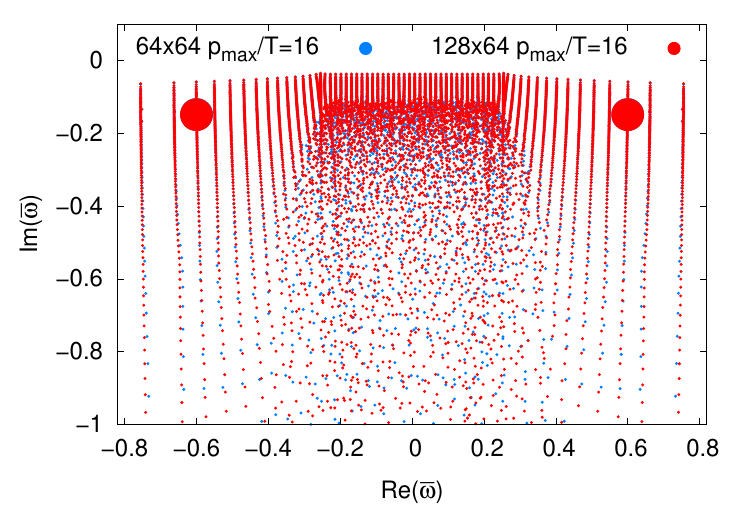}
\includegraphics[width=0.49\textwidth]{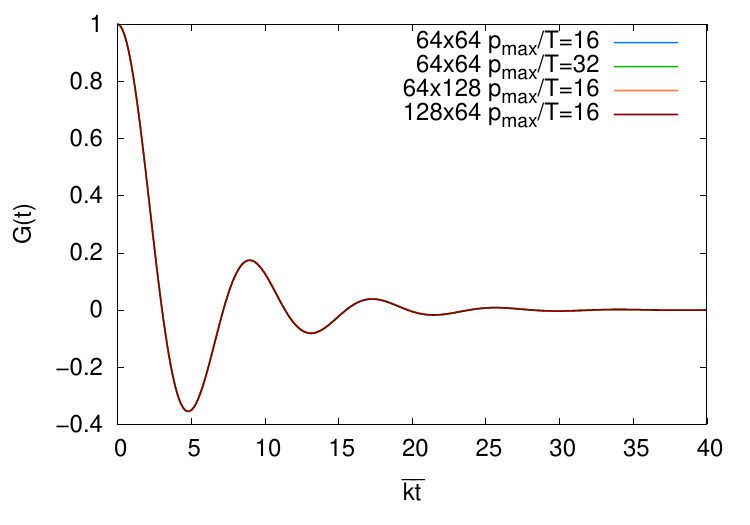}
\caption{Top row and bottom left panel: Eigenvalues of the scalar $\phi^4$ evolution operator in the complex frequency plane for different discretizations. Hydrodynamic modes are highlighted as larger points. Bottom right panel: Real time Green's function $G(\tbar)$ as a function of $\kbar\tbar$ for the same discretizations. All plots have been for a gradient of $\kbar=0.8$.}
\label{fig:discretization}
\end{figure}

Now that we have established that the zero gradient spectrum is a branch cut we will discuss the effects of adding gradients to the Boltzmann equation. Before we discuss the results we should analyze how the discretization plays into it. We denote a certain discretization by $N_p \times\ncos$. For the results later in this work we choose a discretization of $64\times 64$ and $p_{max}/T=16$. In Fig.\ref{fig:discretization} we present the results for three different discretizations for $\kbar=0.8$ in comparison with the final $64\times64$ and $p_{max}/T=16$ discretization.
Eigenvalues of the evolution matrix are plotted in the top row and bottom left panel, with the hydrodynamic eigenvalues highlighted as big dots because they carry the largest individual residue for $\kbar=0.8$.
By increasing the number of angles we see that we get a finer distribution of eigenvalues in the real range of frequencies, as seen in the top right panel. Equivalently we see a refinement in the imaginary frequencies by increasing the number of momenta, as seen in the bottom left panel. Although this yields more eigenvalues the overall picture of the complex plane stays the same, especially the positions of the hydrodynamic poles do not move. When we increase the maximum momentum cut-off, as in the top left panel, we see that we gather more eigenvalues closer to the real axis, as we have already seen for the case of $\kbar=0$. But again this does not change the position of hydrodynamic eigenvalues. Despite the various refinements in the complex frequency plane, the real time Green's functions, plotted in the bottom right panel of Fig.\ref{fig:discretization}, shows the same behavior, where all curves of various discretizations lie on top of each other.  \\ 
This discretization test shows that the discretization of $64\times64$ with $p_{max}/T=16$ is well suited for the study of the complex structure of scalar theory. The further increase in discretization only added eigenvalues with small residues which did not change the complex structure at all and left the eigenvalues with the largest residues untouched. The physical description of observables, represented by the Green's function, remains completely unchanged upon further refinement.

\subsection{Spectrum for finite gradients}
\begin{figure}[ht]
\centering
\includegraphics[width=0.49\textwidth]{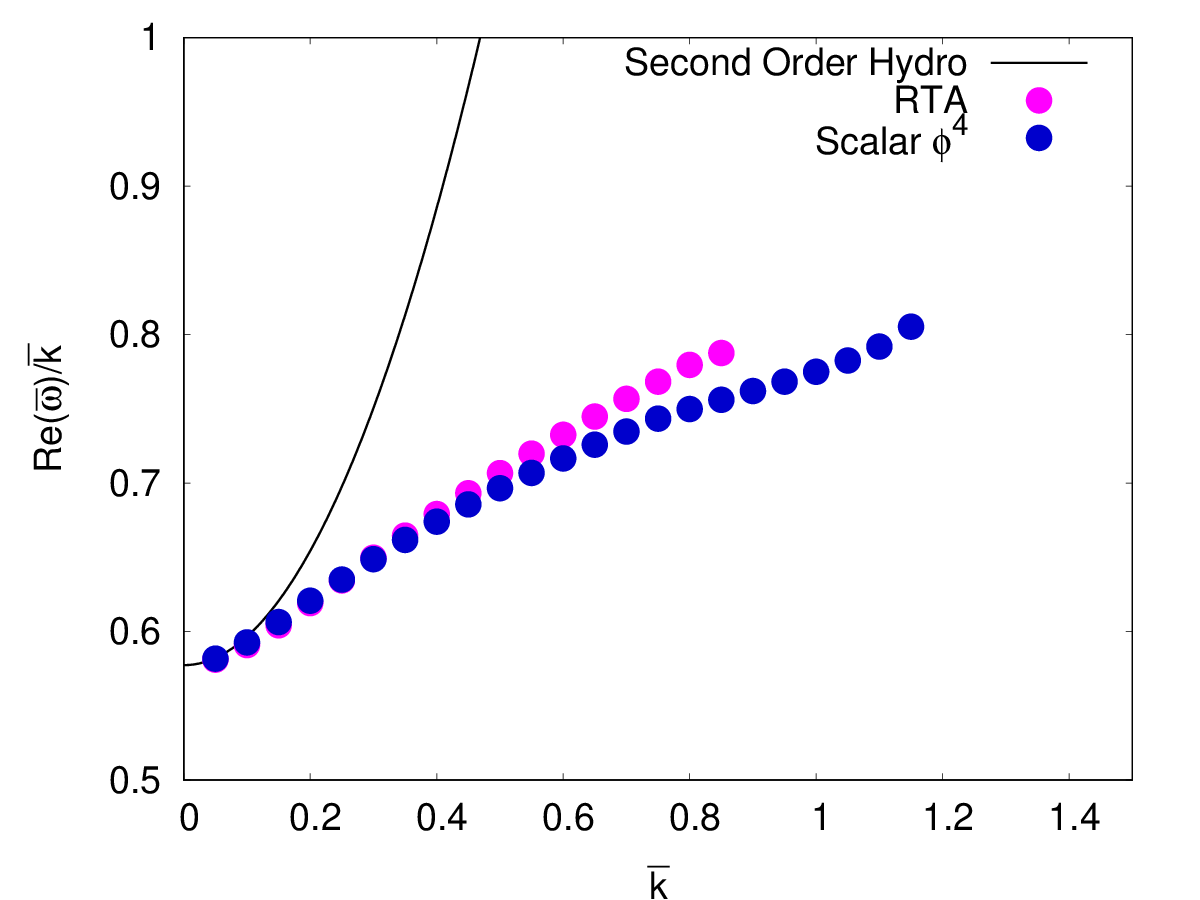}
\includegraphics[width=0.49\textwidth]{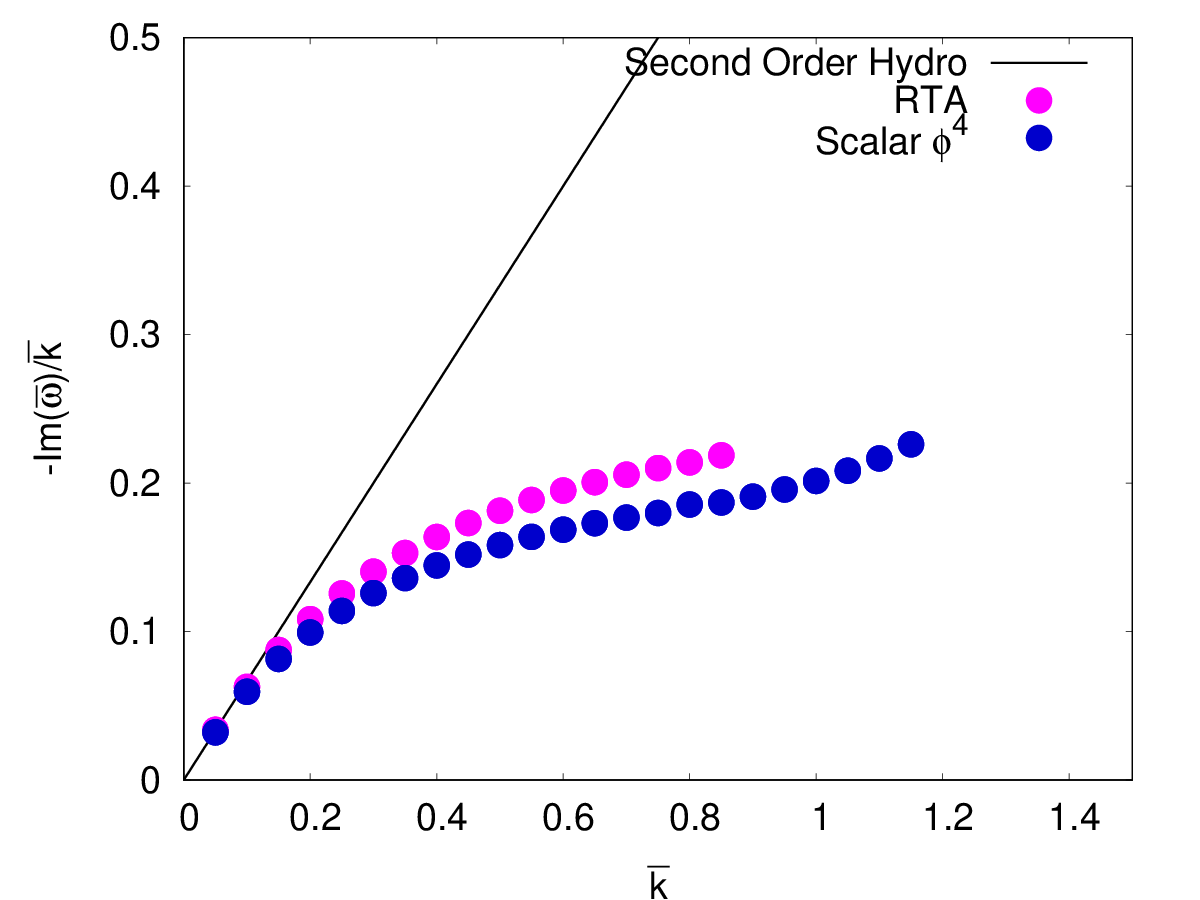}
\caption{The real (left) and imaginary (right) part of the dispersion relation $\Bar{\omega}(\Bar{k})$ of the hydrodynamic sound mode as functions of $\Bar{k}$ in scalar $\phi^4$ theory (blue) and RTA (magenta)}. Black curves show the dispersion relation in second order hydrodynamics in Eq.(\ref{eq:brsss}) for comparison.
\label{fig:dispersion}
\end{figure}

\begin{figure}[H]
\centering
\includegraphics[width=\textwidth]{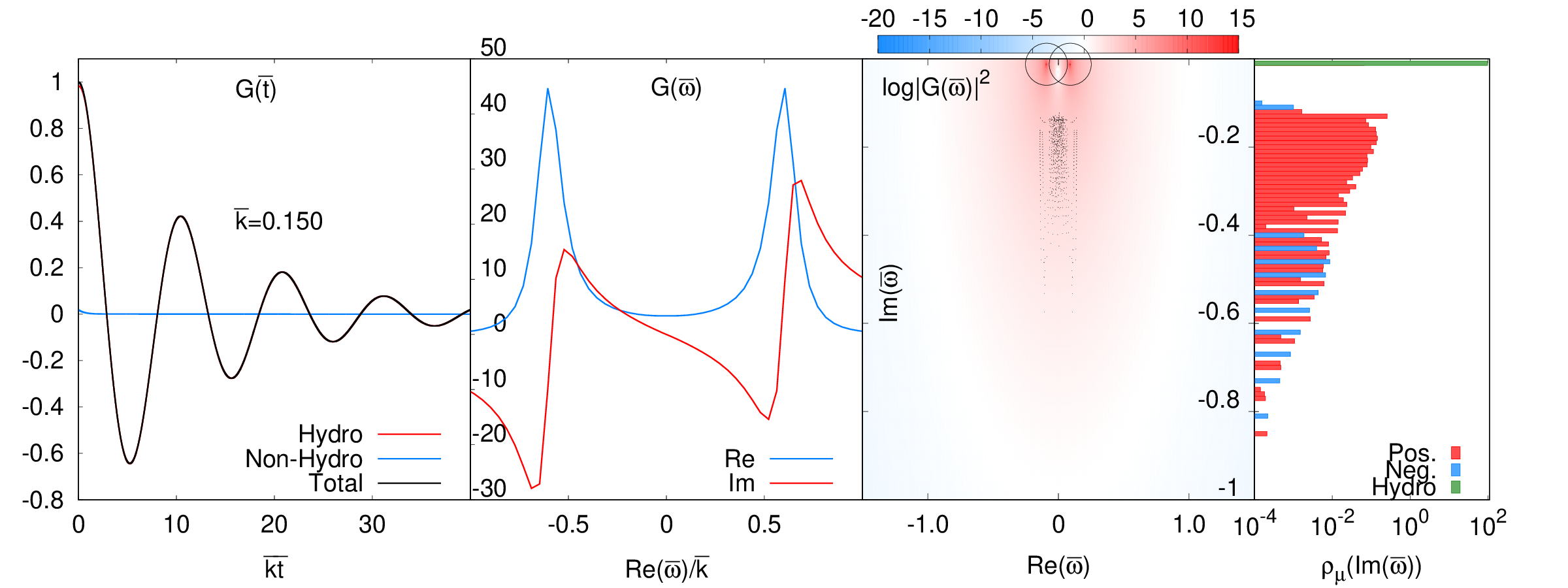}
\includegraphics[width=\textwidth]{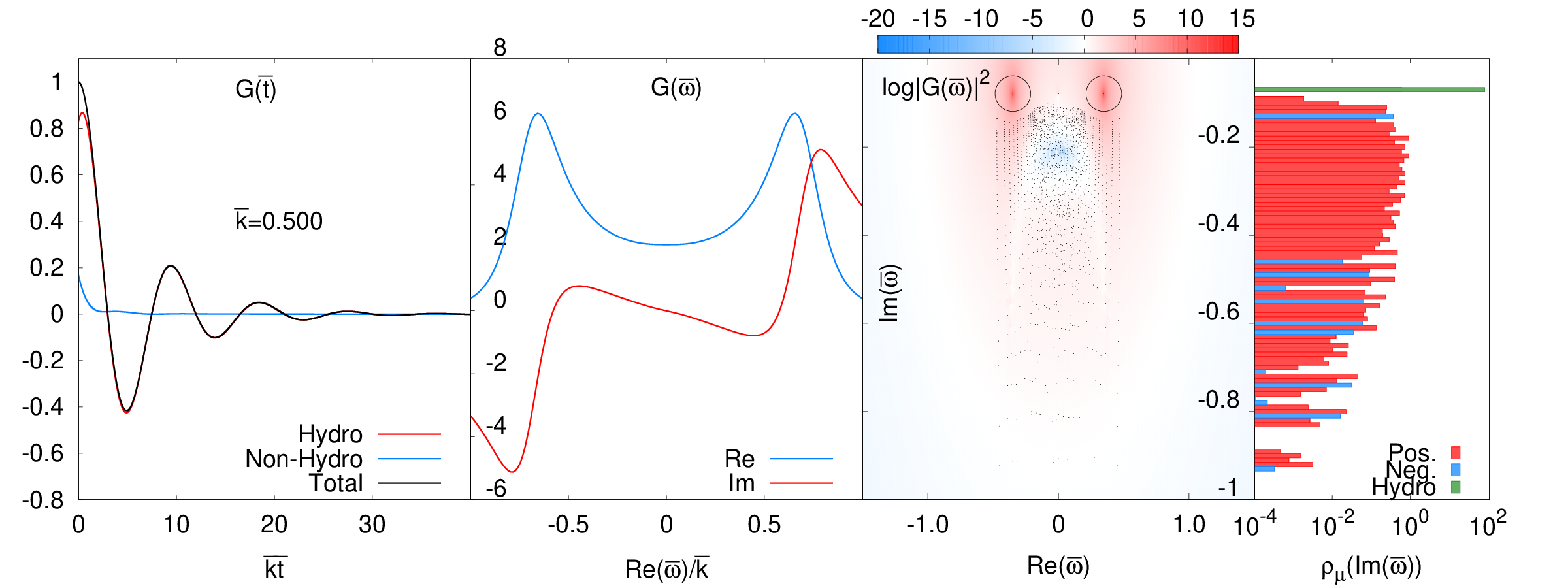}
\includegraphics[width=\textwidth]{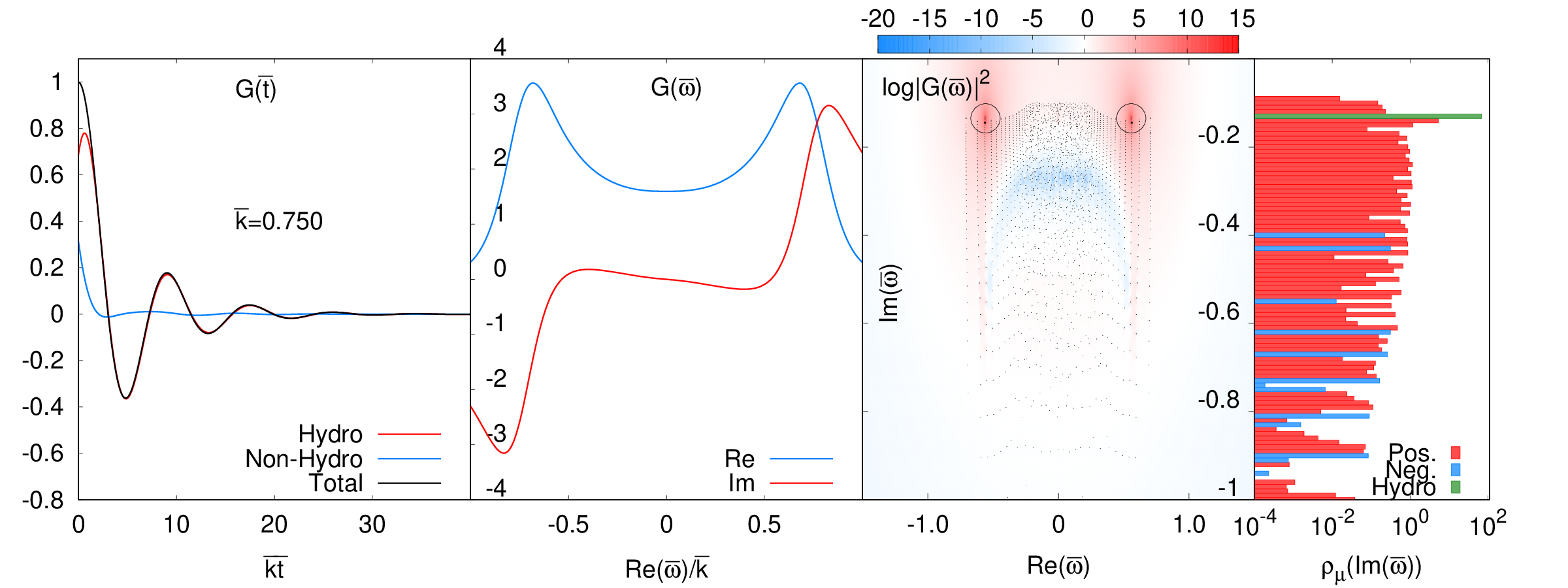}
\caption{From left to right: 1. The real time Green's function $G(\tbar)$ in scalar $\phi^4$ theory decomposed into hydrodynamic and non-hydrodynamic contributions as a function of $\kbar\tbar$. 2. The real frequency Green's function $G(\omegabar)$ split into real and imaginary part as a function of the real frequency $\text{Re}(\omegabar)/\kbar$. 3. The eigenvalues of the evolution operator as black circles in the complex frequency plane. The eigenvalue circle sizes are scaled by their respective residues $\mu_i$. The coloring in the plane is the logarithm of the absolute square of the Green's function $\log(|G(\omegabar)|^2)$ at that position in frequency space. 4. The residue density summed out over the real frequency range and plotted as function of Im$(\omegabar)$, the hydrodynamic mode residue is separate as green bar as comparison. The results in each row are obtained for one gradient $\kbar$.}
\label{fig:finite_gradients1}
\end{figure}

\begin{figure}[H]
\centering
\includegraphics[width=\textwidth]{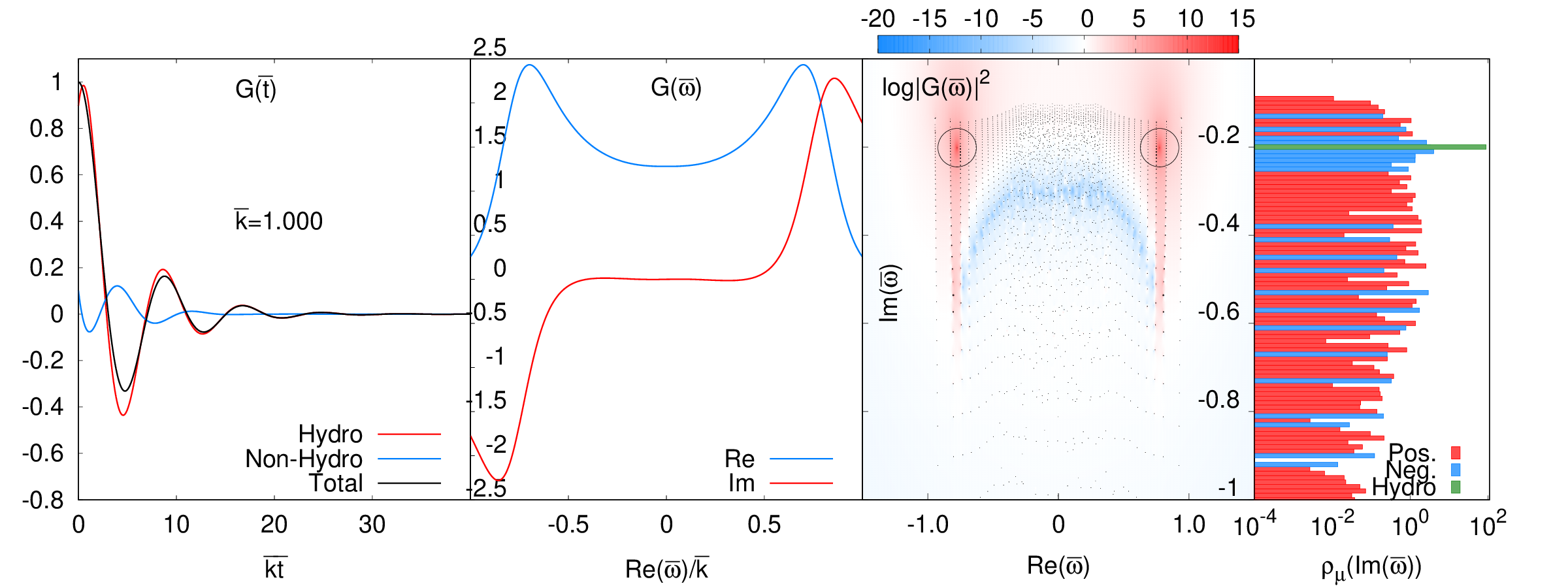}
\includegraphics[width=\textwidth]{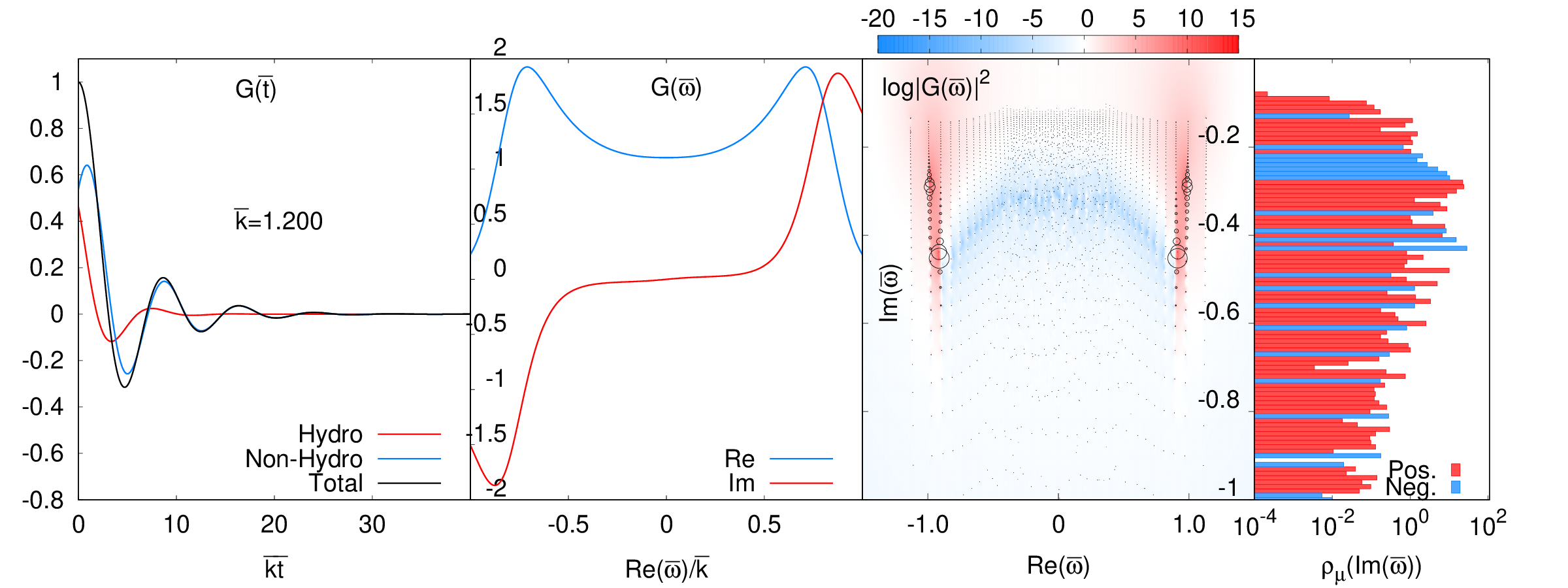}
\includegraphics[width=\textwidth]{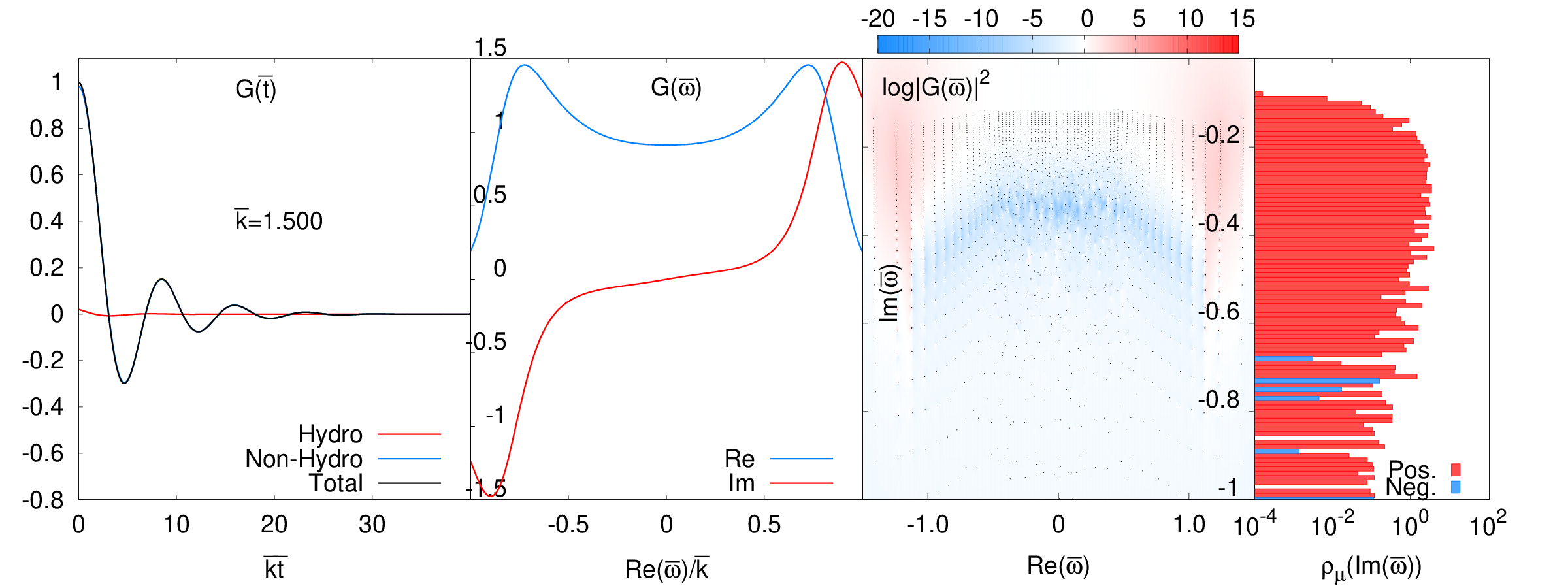}
\caption{From left to right: 1. The real time Green's function $G(\tbar)$ in scalar $\phi^4$ theory decomposed into hydrodynamic and non-hydrodynamic contributions as a function of $\kbar\tbar$. 2. The real frequency Green's function $G(\omegabar)$ split into real and imaginary part as a function of the real frequency $\text{Re}(\omegabar)/\kbar$. 3. The eigenvalues of the evolution operator as black circles in the complex frequency plane. The eigenvalue circle sizes are scaled by their respective residues $\mu_i$. The coloring in the plane is the logarithm of the absolute square of the Green's function $\log(|G(\omegabar)|^2)$ at that position in frequency space. 4. The residue density summed out over the real frequency range and plotted as function of Im$(\omegabar)$, the hydrodynamic mode residue is separate as green bar as comparison. Each row shows the results for a different gradient $\kbar$.}
\label{fig:finite_gradients2}
\end{figure}

When one switches to non zero gradients the appearance of hydrodynamic modes is one of the first things happening for very small $\Bar{k}$. The zero modes observed for $\Bar{k}=0$, responsible for conservation laws, become hydrodynamic modes in the finite $\Bar{k}$ case. All the other modes describe the non-hydrodynamic behavior. \\
Since in our calculation we only receive the location of eigenvalues $\lambda_i$ and their contribution $\mu_i$ to the Green's function, we first need to come up with a definition for what we call hydrodynamic mode and how we isolate it in our data. We define the hydrodynamic modes, where possible (more on that later), as the complex conjugated pair of modes with the largest residue, which is obviously true for small $\Bar{k}$. By identifying these modes for various $\Bar{k}$, we can then obtain the dispersion relation $\Bar{\omega}(\bar{k})$ of hydrodynamic sound modes, which is shown in Fig.~\ref{fig:dispersion} for both scalar theory and RTA as a comparison. 
While at small $\Bar{k}$, the dispersion relations for the real (left) and imaginary (right) part of the sound mode agree with second order viscous hydrodynamics, sizeable deviations start to occur for $\bar{k} \gtrsim 0.15$, and increase with increasing gradient strength, which is in line with the results reported in~\cite{Du:2023bwi}. Beyond $\Bar{k} \approx 1.2$ the residue of the hydrodynamic mode extracted from scalar theory shrinks dramatically, as the mode disappears into a continuum of non-hydrodynamic excitations, and it is no longer meaningful to distinguish this mode from other excitations of the system, which is the reason that the curves in Fig.~\ref{fig:dispersion} terminate at this value of $\bar{k}$. Strikingly, a very similar behavior can also be seen in the Relaxation Time Approximation, where the hydrodynamic modes exhibit essentially the same dispersion relations up to $\kbar\sim 0.9$, where the hydrodynamic modes in RTA disappear behind the branch cut. \\ 
Now that we have established the behavior of the hydrodynamic sound mode, we continue to further analyze the behavior of the Green's functions and in particular investigate the structure and impact of non-hydrodynamic excitations. In order to perform this analysis, we monitor the behavior of the Green's functions in the real-time and real and complex frequency space, while varying the wave-number $\bar{k}$, which characterizes the magnitude of spatial gradients.
Our results are compactly summarized in Figs.~\ref{fig:finite_gradients1} and \ref{fig:finite_gradients2}, where each row shows the behavior of the Green's function for a fixed value of $\kbar$, with increasing $\bar{k}$ from top to bottom. In each row, the left most column shows the behavior of the real time Green's functions $G(\Bar{t})$ (see Eq.(\ref{eq:real_green_decomposed})), along with a decomposition into the contributions from the hydrodynamic sound mode (i.e. the complex conjugated pair of modes with the largest residue) and the non-hydrodynamic modes (i.e. all other), which can be reconstructed by summing only the contributions of the respective modes in Eq.(\ref{eq:real_green_decomposed}). In the second column from the left, we depict the real and imaginary parts of the Green's function $G(\Bar{\omega})$ (see Eq.(\ref{eq:freq_green_decomposed})) as a function of real frequencies $\Bar{\omega}/\Bar{k}$. The behavior in the complex frequency plane is elucidated in the third column of Figs.~\ref{fig:finite_gradients1} and \ref{fig:finite_gradients2}, where black circles show the location of individual eigenvalues calculated from the matrix, which are scaled in size by their contribution $\mu_i$ to the Green's function, while the color code in these plots corresponds to the logarithm of the absolute square of the Green's function.\footnote{Note that one does not see every eigenvalue here because some have such a small contribution that their sizes are scaled to be invisible, so one only sees eigenvalues that have at least some influence on the Green's function.} Finally, the right most plot is a histogram showing the summed contributions $\rho_\mu(\omegabar)$ from non-hydrodynamic modes as defined in Eq.(\ref{eq:residue_density}).
Since there can be both positive and negative contributions, we distinguish them by a color coding, where positive contributions are plotted in red and negative contributions in blue. Contributions of the non-hydrodynamic modes are further compared to the contribution of the the hydrodynamic sound mode, which is indicated by a  green bar. \\
When considering the behavior of the real-time Green's function $G(t)$ (left column), we find that for small gradients $\Bar{k} \ll 1$ the evolution is almost purely hydrodynamic, meaning that only the complex conjugated pair of modes with the largest residue play a significant role for the time evolution of the system. With increasing $\Bar{k}$, the contribution of non-hydrodynamic modes becomes visible at early times, but exhibits a much faster decay than the hydrodynamic contribution. When $\Bar{k}\sim 1$, the contributions from hydrodynamic and non-hydrodynamic modes become of comparable size, until for $\bar{k} \approx 1.2$ the contribution of the hydrodynamic mode begins to disappear as non-hydrodynamic modes start to dominate the behavior of the real-time Green's function. By $\bar{k}=1.5$, corresponding to the largest value shown in Fig.\ref{fig:finite_gradients2}, it is then no longer possible to identify a clear signature of a hydrodynamic mode in the system.

Next we consider the Green's function $G(\bar{w})$ as a function of real frequency $\bar{\omega}$ (second left column), which for small gradients $(\Bar{k} \ll 1)$ features two distinct peaks in its real part and at the same position two steep inflection points in the imaginary part. Clearly, these features are a sign of the contribution of two very distinct regions in the complex frequency plane, in this case the hydrodynamic poles, which for $\Bar{k}\ll 1$ are located close to real frequency axis. By increasing $\Bar{k}$ the influence of the hydrodynamic poles diminishes, as their residues shrink and they move away from the real-frequency axis, leading to a broadening of peaks and inflections. Nevertheless, both the real-time and the real-frequency Green's functions
exhibit a remarkably smooth behavior with increasing $\Bar{k}$, and do not immediately indicate a transition from a hydrodynamic to a non-hydrodynamic regime. 

Now, if we consider the behavior of the Green's function in the complex frequency plane (right two columns), for small $\Bar{k}\ll 1$ we can clearly see the dominance of the hydrodynamic modes, which correspond to simple poles in the complex frequency plane with large residues. However, in addition to the hydrodynamic poles, additional singularities indicated by black circles appear throughout the region where $-\Bar{k}<\text{Re}(\Bar{\omega}) <\Bar{k}$ and $\text{Im}(\Bar{\omega})<0$, which most likely signals the presence of an entire region of non-analyticity as discussed in \cite{Kurkela:2017xis} for the momentum dependent relaxation time approximation. Despite the fact that the singularities cover a large region in the complex frequency plane, it appears that the dominant contribution to the Green's function of the energy-momentum tensor seems to be located around $\text{Im}(\Bar{\omega})\approx -0.2$, as can be seen from the residue density in the right most panel, which reminiscent of the $\bar{k}=0$ spectrum in Fig.~\ref{fig:scalar_zerok} is also strongly peaked around a single imaginary part of the frequency. Hence, the typical time scale for the relaxation of contributions of non-hydrodynamic modes to the energy-momentum tensor is still given by the inverse of this characteristic frequency $\sim 1/\text{Im}(\bar{\omega})\approx 5$, which is in line with the value of the second order transport coefficient $\bar{\tau}_{\pi} \approx 6.1 $~\cite{York:2008rr}. 
Due to this rather strong peak in the residue density, the behavior of the Green's function in scalar $\phi^{4}$ theory is actually not to different from the behavior in the conformal relaxation time approximation (see also ~\cite{Du:2023bwi}) where -- instead of a spread out region in the complex frequency plane -- the non-hydrodynamic contributions originates from a single branch-cut located at $\text{Im}(\Bar{\omega})=-1/\bar{\tau}_R=0.2$.

When increasing $\Bar{k}$, the hydrodynamic sound mode moves further out into the complex frequency plane and its residue decreases; at the same time the overall contribution of the non-hydrodynamic modes increases, without any dramatic changes in the spectrum of the residue density of the contributing modes. Even though for $\Bar{k}=0.75$, the characteristic time scales for the decay of hydrodynamic and non-hydrodynamic modes is comparable, the contribution of the hydrodynamic mode still stands out. With higher $\Bar{k}$ the complex conjugated pair of hydrodynamic modes slowly dives deeper and deeper into the non-analytic continuum until around $\Bar{k}=1.2$, where it gets absorbed into it and is no longer distinguishable.  Eventually, for large $\Bar{k}$ the remnant is a large non-analytic region with fairly uniform impact on the Green's function. Nevertheless, one can still see which parts contribute more via the color coding of the Green's function, showing that the largest contributors are located at the edge of the region, towards $\text{Re}(\Bar{\omega})=\pm k$.

% At the same time we observe in the real time Green's functions that the non-hydrodynamic part dominates the hydrodynamic part. Up to this point the definition of our sound poles was that they are the two largest modes, but in a situation where all modes have a similar significance we can no longer guarantee that they are not overshadowed by two random non-hydrodynamic modes. Revisiting the real time evolution one has to add that the hydrodynamic part from here on just refers to the two largest modes, so the hydrodynamic evolution for larger $\Bar{k}$ is to be taken with a grain of salt. If we now go to even higher $\Bar{k}$ the area fans out further and the amount of eigenvalues that contribute significantly increase while also decreasing their average size. At this point the Green's function is determined by the whole region. \\ 
% This behavior is mirrored in the $k$ dependence of the residues. In the beginning only the sound modes and the region around $-1/\tau_R$ is important, note the logarithmic scale here. increasing $\Bar{k}$ shrinks the contribution by sound modes while broadening and increasing the influence of the non-analytic region. For $\Bar{k}=1.2$ and going upwards we dismiss the sound mode because it is no longer reasonable to link certain modes to Hydrodynamics.

\section{Conclusions \& Outlook} \label{seq:conclusion}
We developed a new method to numerically calculate linear response functions of the energy-momentum tensor in kinetic theory. By carefully discretizing the Boltzmann equation on a momentum grid, this method allows to extract eigenvalues and eigenfunctions of the evolution operator to access the behavior of the Green's function in the complex frequency plane, without the need to perform a numerical Laplace transform. The formalism was tested within the well studied conformal Relaxation Time Approximation and could reproduce the known analytic structure of its Green's functions, including the non-hydrodynamic cut in the lower complex plane. 

Subsequently, we explored for the first time the analytic structure of the Green's function of the energy momentum tensor in the sound channel, for a scalar field theory with quartic self-interaction. We find that for small gradients the response of the system is dominated by hydrodynamic modes, which are embedded as single poles within a continuum of non-hydrodynamic excitations corresponding to a non-analytic region in the complex frequency plane that extends arbitrarily close to the real frequency axis, albeit with exponentially suppressed contributions. Nevertheless, for sufficiently small wave numbers, the contribution of non-hydrodynamic excitations to physical observables is strongly peaked around a characteristic value of the imaginary part of the frequency Im$(\omegabar)\sim -\bar{\tau}^{-1}_{\pi}$. With increasing gradient strength the non-hydrodynamic modes gain more and more influence in the response, until around $\bar{k} \gtrsim 1.2$ the hydrodynamic modes disappear into the continuum and the response of the energy-momentum tensor is entirely determined by non-hydrodynamic excitations.

Based on our analysis, it is thus conceivable that generalized hydrodynamic theories accounting for higher gradient corrections to properly capture the dispersion relations can provide a valid effective description of kinetic theory for sufficiently small gradients $\bar{k} \lesssim 1$, where Green's functions of the energy-momentum tensor in the sound channel are dominated by the hydrodynamic modes and a subset of non-hydrodynamic excitations with $\text{Im}(\Bar{\omega})\approx -1/\bar{\tau}_{\pi}$ which could be captured by effective non-hydrodynamic modes. However, in the presence of large gradients $\Bar{k} \gtrsim 1$, where a continuum of non-hydrodynamic modes contributes to the Green's function, thus reflecting the underlying microscopic dynamics, it is not clear how hydrodynamics can provide a meaningful description of the dynamics of the system. Since our analysis of the modes is of numerical nature, it is hard to determine the precise location, where the hydrodynamic modes are completely hidden behind the non-hydrodynamic continuum. Moreover, our results indicate that the transition between the two regimes is not particularly sharp, but rather a smooth transition from one to another where hydrodynamic and non-hydrodynamic modes exchange their relative weights. Nevertheless, the location of the threshold $\kbar_{c} \sim 1$ is in rather good agreement with the value $\kbar_{c} \sim 0.9$ previously obtained in RTA~\cite{Romatschke:2015gic,Romatschke:2016hle}, which is also reproduced by our numerical RTA results. By converting this estimate into coordinate space, one obtains a critical length scale $l_c\sim1/k_{c} \approx 0.16~ {\rm fm}\left(\frac{200 {\rm MeV}}{T}\right)\left(\frac{\eta/s}{0.16}\right)$ for typical values of the temperature and transport properties of the QGP. Below this length scale hydrodynamic modes become suppressed and non-hydrodynamic excitations govern the dynamics. The critical length scale is extremely small and should be seen as a lower bound for the applicability of hydrodynamics.

Evidently this study provides a first step to an analogous calculation in QCD kinetic theory, where on general grounds one also expect a non-analytic structure that is far more complicated than just poles and cuts \cite{Moore:2018mma,Kurkela:2017xis,Romatschke:2017ejr}. Beyond the extension to QCD kinetic theory, it would also be interesting to extend the present study in scalar field, from a kinetic description to a genuine QFT treatment, which for weakly coupled theories could be achieved based on $n$-particle irreducible effective action techniques~\cite{Berges:2004yj}.

\section{Acknowledgement}
We thank Travis Dore, Xiaojian Du, Guy D. Moore, Philip Plaschke, Paul Romatschke, and Ismail Soudi for their valuable discussions. This work is supported by the Deutsche Forschungsgemeinschaft (DFG) under grant CRC-TR 211
“Strong-interaction matter under extreme conditions” project no. 315477589-TRR 211. The authors gratefully acknowledge computing time provided by the Paderborn Center for Parallel
Computing (PC2).

\appendix{}

\section{Analytical Calculation of RTA Green's Functions} \label{appendix:rta_analytic}
In this part we will further discuss the analytical solution of the Boltzmann equation. We recall that the distribution function solving the perturbed equation is
\begin{align}
\delta f_k(\bfp,\omega)=\frac{\delta f_{k,\rm eq}(\bfp,\omega)+\delta f_k(\bfp,t=0)\tau_R}{1+ik\tau_R\cos\theta-i\omega\tau_R} \ . \label{eq:rta_distribution}
\end{align}
To keep in mind the perturbed equilibrium distribution is
\begin{align}
\delta f_{k,\rm eq}(\bfp,t)=\frac{p}{T}f_{\rm eq}^2\left(\frac{\delta T_k}{T}-\delta u^\mu_k v_\mu\right)e^{p/T} \ .
\end{align}
Using Landau matching conditions for the energy momentum tensor (Eq.(\ref{eq:energy_momentum_tensor})) we can express the temperature and velocity perturbation as moments of the distribution function 
\begin{align}
\frac{\delta T_k}{T}&=\frac{1}{4}\frac{\delta e_k}{e}=\frac{1}{4e}\int \frac{d^3p}{(2\pi)^3}p \delta f_k \ ,
\\
\delta u^\mu_k &= \frac{\delta T^{0\mu}_k}{e+P}=\frac{3}{4e}\int \frac{d^3p}{(2\pi)^3}p^\mu \delta f_k \ .
\end{align}
Plugging in everything we know into Eq.(\ref{eq:energy_momentum_tensor}) we receive
\begin{align}
\delta e_k=\delta T^{00}_k&=\int_0^\infty \frac{dp}{2\pi^2T}p^4 e^{p/T} f_{\rm eq}^2\int_0^{2\pi}\frac{d\phi}{4\pi}\int_{-1}^1 d\cos\theta~\frac{\frac{1}{4}\frac{\delta e_k}{e}+\frac{3}{4}\frac{\delta T^{03}_k\cos\theta}{e}+\frac{\tau_R}{4}\frac{\delta e_0}{e}}{1+ik\tau_R\cos\theta-i\omega \tau_R} \\
&=\int_{-1}^1d\cos\theta~\frac{\frac{1}{2}\delta e_k+\frac{3}{2}\delta T^{03}\cos\theta+\frac{1}{2}\tau_R \delta e_0}{1+ik\tau_R\cos\theta-i\omega\tau_R} .
\end{align}
And for the off-diagonal component
\begin{align}
\delta T^{03}_k=\int_{-1}^1 d\cos\theta\frac{\frac{1}{2}\delta e_k+\frac{3}{2}\delta T^{03}\cos\theta +\frac{1}{2}\tau_R\delta e_0}{1+ik\tau_R\cos\theta-i\omega\tau_R}\cos\theta.
\end{align}
Together they form a system of linear equations
\begin{align}
\delta e_k &= a \delta e_k + b \delta T^{03}_k +c \ , \\
\delta T^{03}_k &= d \delta e_k + g\delta T^{03}_k + h \ ,
\end{align}
where each individual coefficient is the integral of one part of the sums. The solution to this system of equations is
\begin{align}
\delta e_k = \frac{c-cg+bh}{1-a-bd-g+ag} \ , \quad \delta T^{03}_k=\frac{cd+h-ah}{1-a-bd-g+ag} \ . \label{eq:appendix:lsoe}
\end{align}
The coefficients evaluate to
\begin{align}
a&=\frac{iL}{2k\tau_R} \ , \quad b=i\frac{3(i+\omega\tau_R)L-6k\tau_R}{2k^2\tau_R^2} \ , \quad c=\frac{i\delta e_0}{2k}L \ , \\
d&=-i\frac{2k\tau_R-(i+\omega\tau_R)L}{2k^2\tau_R^2} \ , \quad g=-\frac{3i(i+\omega\tau_R)(2k\tau_R-(i+\omega\tau_R)L)}{2k^3\tau_R^3} \ , \\
h&=-i\delta e_0\frac{2k\tau_R-(i+\omega\tau_R)L}{2k^2\tau_R} \ ,
\end{align}
where we used $L=\log(\frac{1-i\tau_R(k+\omega)}{1+i\tau_R(k-\omega)})$. Plugging these coefficients into Eq.(\ref{eq:appendix:lsoe}) yields the energy momentum tensor components as
\begin{align}
\frac{\delta e_k}{\delta e_0}=\frac{-(6ik\tau_R+(3+k^2\tau_R^2-3i\omega\tau_R)L)}{(k^2\tau_RL+2ik^3\tau_R^2-6k\tau_R\omega+3\omega(i+\omega\tau_R)L)} \ , \\
\frac{\delta T^{03}_k}{\delta e_0}=\frac{ik\tau_R(-2k\tau_R+(i+\omega\tau_R)L)}{(k^2\tau_RL+2ik^3\tau_R^2-6k\tau_R\omega+3\omega(i+\omega\tau_R)L)} \ .
\end{align}

\section{Discretization of RTA Collision Operator in Wedge Moments} \label{appendix:rta_numerics}

We start with the perturbed Boltzmann Equation in RTA (Eq.(\ref{eq:perturbed_rta_timeevo}))
\begin{align}
\partial_t \delta f_k(\bfp,t) + ik\cos\theta~ \delta f_k(\bfp,t)=\frac{1}{\tau_R} e^{p/T} f_{\rm eq}^2(p)\left(\frac{p}{T}\frac{\delta T_k}{T}-\frac{\delta u^\mu_k p_\mu}{T}\right)-\frac{1}{\tau_R}\delta f_k (\bfp,t)\ .
\end{align}
The right hand side will be expanded in moment space as the left side only contains the time derivative and expansive term, which was already discussed generally. The moments of the collision kernel on the right side are given as
\begin{align}
\delta C_i=-\frac{1}{\tau_R}N_i+\frac{1}{\tau_R}\int \frac{d^3p}{(2\pi)^3}w_\ipe(p) w_\itheta(\cos\theta)e^pf_{\rm eq}^2p\left(\delta T^{00}+3\delta T^{0m}v_m\right)\frac{1}{4e} \ .
\end{align}
We recognize that the energy momentum tensor components can be directly decomposed into wedge moments via Eq.(\ref{macroreconstruction}). Integrals with $m\neq3$ vanish. When we take the functional derivative with respect to $N_j$ only the $j-$th contributions to the energy momentum components stay. The matrix elements then are
\begin{align}
C_{ij}=-\frac{\delta_{ij}}{\tau_R}+\frac{1}{\tau_R}\frac{1}{4e}\int \frac{d^3p}{(2\pi)^3}p w_\ipe(p)w_\itheta(\cos\theta)e^pf_{\rm eq}^2(p_\jpe+3p_\jpe\cos\theta_\jtheta\cos\theta) \ ,
\end{align}
which can finally be simplified to the form
\begin{align}
C_{ij}=-\frac{\delta_{ij}}{\tau_R}+\frac{1}{\tau_R}\frac{p_\jpe}{4e}\int_0^\infty \frac{dp}{(2\pi)^2}p^3w_\ipe(p)e^pf_{\rm eq}^2\int_{-1}^1d\cos\theta w_\itheta(1+3\cos\theta_\jtheta\cos\theta) \ .
\end{align}
The cosine integral is calculated analytically and only has different values for combinations of $i$ and $j$, which are tabulated. The $p$ integral is calculated numerically and tabulated for all possible $i$.

\bibliographystyle{JHEP}

\bibliography{ref.bib}

\end{document}